\documentclass[sigconf, nonacm]{acmart}

\newcommand\vldbdoi{XX.XX/XXX.XX}
\newcommand\vldbpages{XXX-XXX}
\newcommand\vldbvolume{19}
\newcommand\vldbissue{8}
\newcommand\vldbyear{2026}
\newcommand\vldbauthors{\authors}
\newcommand\vldbtitle{\shorttitle} 
\newcommand\vldbavailabilityurl{https://github.com/dmemsys/CIDER}
\newcommand\vldbpagestyle{empty}

\usepackage{tikz}
\usetikzlibrary{positioning}

\usepackage{array}
\usepackage{makecell}
\usepackage{amsmath}
\usepackage{xspace}
\usepackage{enumitem}
\usepackage{wrapfig}
\usepackage{float}
\usepackage[font=small]{caption}
\usepackage{subcaption}
\usepackage[utf8]{inputenc}
\usepackage{fontawesome}
\usepackage{libertine}
\usepackage{color}
\definecolor{darkgreen}{RGB}{0, 102, 5}
\definecolor{mygraygreen}{RGB}{64, 128, 128}
\usepackage[]{collab}
\usepackage{soul}
\usepackage{booktabs}
\usepackage{threeparttable}
\usepackage[ruled,vlined,linesnumbered]{algorithm2e}
\usepackage{xcolor}
\usepackage[export]{adjustbox}
\usepackage[most]{tcolorbox}

\newcommand{\DMCS}{CIDER\xspace}

\newcommand{\ie}{{\em i.e.},\xspace}
\newcommand{\eg}{{\em e.g.},\xspace}

\usepackage{titlesec}
\titlespacing*{\section}{0pt}{4pt plus 2pt minus 1pt}{2pt plus 1pt minus 1pt}
\titlespacing*{\subsection}{0pt}{3pt plus 1pt minus 1pt}{1pt plus 1pt minus 1pt}
\titlespacing*{\subsubsection}{0pt}{2pt plus 1pt minus 1pt}{1pt plus 1pt minus 1pt}
\titleformat{\subsubsection}[runin]
  {\normalfont\normalsize\itshape}
  {\thesubsubsection}{3mm plus 1mm minus 1mm}{}[.\hspace*{0.6mm}]

\begin{document}

\date{}

\title{\DMCS: Boosting Memory-Disaggregated Key-Value Stores with Pessimistic Synchronization}

\settopmatter{authorsperrow=3}
\author{Yuxuan Du}
\affiliation{%
\normalsize
College of Computer Science and Artificial Intelligence, Fudan~University
}
\affiliation{
\normalsize
National Key Laboratory of Parallel and Distributed Computing, China}
\email{duyx23@m.fudan.edu.cn}

\author{Xuchuan Luo}
\affiliation{%
\normalsize
College of Computer Science and Artificial Intelligence, Fudan~University
}
\affiliation{
\normalsize
National Key Laboratory of Parallel and Distributed Computing, China}
\email{xcluo23@m.fudan.edu.cn}

\author{Xin Wang}
\affiliation{%
\normalsize
College of Computer Science and Artificial Intelligence, Fudan~University
}
\email{xinw@fudan.edu.cn}

\author{Yangfan Zhou}
\affiliation{%
\normalsize
College of Computer Science and Artificial Intelligence, Fudan~University
}
\affiliation{
\normalsize
National Key Laboratory of Parallel and Distributed Computing, China}
\email{zyf@fudan.edu.cn}

\author{Jiacheng Shen}
\affiliation{%
\normalsize
Duke Kunshan University
}
\email{jc.shen@dukekunshan.edu.cn}

\begin{abstract}
Memory-disaggregated key-value (KV) stores suffer from a severe performance bottleneck due to their I/O redundancy issues.
A huge amount of redundant I/Os are generated when synchronizing concurrent data accesses, making the limited network between the compute and memory pools of DM a performance bottleneck.
We identify the root cause for the redundant I/O lies in the mismatch between the optimistic synchronization of existing memory-disaggregated KV stores and the highly concurrent workloads on DM.
In this paper, we propose to boost memory-disaggregated KV stores with pessimistic synchronization.
We propose \DMCS, a compute-side I/O optimization framework, to verify our idea.
\DMCS adopts a \textit{global write-combining} technique to further reduce cross-node redundant I/Os.
A \textit{contention-aware synchronization} scheme is designed to improve the performance of pessimistic synchronization under low contention scenarios.
Experimental results show that \DMCS effectively improves the throughput of state-of-the-art memory-disaggregated KV stores by up to $6.6\times$ under the YCSB benchmark.
\end{abstract}

\maketitle

\pagestyle{\vldbpagestyle}
\begingroup\small\noindent\raggedright\textbf{PVLDB Reference Format:}\\
\vldbauthors. \vldbtitle. PVLDB, \vldbvolume(\vldbissue): \vldbpages, \vldbyear.\\
\href{https://doi.org/\vldbdoi}{doi:\vldbdoi}
\endgroup
\begingroup
\renewcommand\thefootnote{}\footnote{\noindent
This work is licensed under the Creative Commons BY-NC-ND 4.0 International License. Visit \url{https://creativecommons.org/licenses/by-nc-nd/4.0/} to view a copy of this license. For any use beyond those covered by this license, obtain permission by emailing \href{mailto:info@vldb.org}{info@vldb.org}. Copyright is held by the owner/author(s). Publication rights licensed to the VLDB Endowment. \\
\raggedright Proceedings of the VLDB Endowment, Vol. \vldbvolume, No. \vldbissue\ %
ISSN 2150-8097. \\
\href{https://doi.org/\vldbdoi}{doi:\vldbdoi} \\
}\addtocounter{footnote}{-1}\endgroup

\ifdefempty{\vldbavailabilityurl}{}{
\vspace{.3cm}
\begingroup\small\noindent\raggedright\textbf{PVLDB Artifact Availability:}\\
The source code, data, and/or other artifacts have been made available at \url{\vldbavailabilityurl}.
\endgroup
}

\section{Introduction}
Memory-disaggregated key-value (KV) stores, \ie KV stores on the disaggregated memory (DM) architecture, are widely discussed in both academia and industry ~\cite{luo2023smart,zuo2021race,wang2022sherman,luo2024chime,shen2023ditto,shen2023fusee,lee2023dinomo}.
DM decouples CPU and memory resources from individual monolithic servers into independent compute and memory pools.
Compute nodes in the compute pool access and modify data in the memory pool with high-performance networking, \eg remote direct memory access (RDMA)~\cite{infiniband} and compute express link (CXL)~\cite{cxl}.
Compared with traditional distributed KV stores, memory-disaggregated KV stores can achieve better elasticity and resource efficiency due to their enhanced flexibility in resource management.

To ensure data consistency when client threads\footnote{In the rest of this paper, we use clients as an abbreviation for client threads.} concurrently access and modify data in the memory pool, it is necessary that clients can efficiently synchronize their memory pool operations. Unfortunately, achieving efficient synchronization poses a significant challenge for memory-disaggregated KV stores, particularly under conditions of high skew and substantial concurrency.
In general, two approaches are adopted to synchronize concurrent data accesses, \ie lock-free optimistic synchronization and lock-based pessimistic synchronization.
Existing memory-disaggregated KV stores tend to employ optimistic synchronization schemes to avoid the high lock maintenance overhead~\cite{zuo2021race,chen2020clevel,min2024sephash}.
However, optimistic synchronization approaches suffer from poor performance due to their severe I/O redundancy issues.
Extensive redundant I/Os are generated since optimistic synchronization handles conflicting data accesses and modifications by iteratively retrying conflicting operations.
The redundant I/Os quickly saturate the limited network bandwidth and IOPS of the memory pool, resulting in severe performance degradation.
According to our preliminary experiments, existing memory-disaggregated KV stores suffer from more than $2.7\times$ slowdown due to the wasted IOPS and bandwidth when synchronizing concurrent data accesses.

The primary cause of the I/O redundancy lies in the inadequacy of optimistic synchronization under the highly contended real-world workloads on DM. 
In this paper, we propose to enhance the performance of memory-disaggregated KV stores with lock-based pessimistic synchronization.
Nonetheless, two significant challenges must be overcome to achieve high performance.

\textbf{\textit{1) Inter-node redundant data modifications limit system throughput.}}
While pessimistic synchronization effectively reduces redundant I/Os during data synchronization, it falls short in addressing the I/O redundancy issues during concurrent data modifications. 
Specifically, when multiple clients concurrently modify the same data, the data modified by one client is frequently overwritten by others, causing redundant data modifications. 
Existing approaches mitigate this issue by proposing a local write combining scheme~\cite{luo2023smart,luo2024chime} that consolidates redundant data modifications among multiple clients within their corresponding compute nodes. 
However, this approach only mitigates intra-node redundant data modifications, leaving inter-node I/O redundancy unresolved.

\textbf{\textit{2) High lock overhead under low contention workloads.}}
Although lock-based pessimistic synchronization performs well in highly contended scenarios, its efficiency declines under workloads with low contention.
Specifically, locks have to be maintained in the memory pool to coordinate clients from multiple compute nodes.
Additional remote memory accesses are required to acquire and release locks on the critical path of each data access operation.
Operation latency significantly increases due to the high remote memory access overhead, \ie an order of magnitude higher than local memory accesses.
System throughputs are thus compromised due to the higher per-operation I/O overhead.

We design \DMCS, a compute-side I/O optimization framework, to enhance the performance of memory-disaggregated KV stores with pessimistic synchronization.
\DMCS adopts a distributed \textbf{M}ellor-\textbf{C}rummey-\textbf{S}cott (MCS) lock, the state-of-the-art lock mechanism designed for DM~\cite{gao2025shiftlock}, to implement efficient pessimistic synchronization.
Over the MCS lock, we further propose two techniques to address the above two challenges.
First, to eliminate extensive cross-node redundant data modifications, \DMCS employs a \textit{global write-combining} technique.
It organizes redundant and concurrent data modifications in a global queue and executes them through a single consolidated data modification.
Second, to improve system performance under low contention scenarios, \DMCS designs a \textit{contention-aware synchronization} scheme.
It dynamically identifies the contention level of the current workload on the client side and allows clients to dynamically switch between optimistic and pessimistic synchronization modes.

We implement \DMCS from scratch and evaluate it with both micro-benchmarks and end-to-end evaluation under YCSB workloads~\cite{cooper2010benchmarking}.
For the micro-benchmark, we design a minimalistic object storage (termed as a pointer array), to quantify the pure performance advancement brought about by \DMCS.
For the end-to-end evaluations, we integrate \DMCS into RACE~\cite{zuo2021race} and SMART~\cite{luo2023smart}, two state-of-the-art memory-disaggregated KV stores that adopt hash- and tree-based indexes, respectively, to show the overall performance boost.
Our evaluation results show that \DMCS outperforms the state-of-the-art optimistic and pessimistic synchronization schemes on DM under our micro-benchmark by up to $6.7\times$ and $2.0\times$ in throughput, respectively.
Furthermore, by integrating \DMCS, RACE, and SMART achieve up to $5.1\times$, $6.6\times$ higher throughput and $12.4\times$, $13.8\times$ lower P99 latency under the write-intensive workload, respectively.


The contributions of this paper can be summarized in the following three aspects:
\begin{itemize}[noitemsep, topsep=0pt, parsep=0pt, partopsep=0pt]
    \item We identify the performance issue with optimistic synchronization in existing memory-disaggregated KV stores with thorough experiments.
    \item We propose the idea of enhancing memory-disaggregated KV stores with lock-based pessimistic synchronization and design \DMCS to address the challenges of achieving efficient pessimistic data synchronization on DM.
    \item We implement \DMCS from scratch and evaluate it with extensive experiments. Our evaluation results show that \DMCS boosts the throughput of state-of-the-art memory-disaggregated KV stores by up to $5.1\times$ under the write-intensive workload.
\end{itemize}

\section{Background and Motivations}\label{sec:background-and-motivations}
In this section, we first introduce the disaggregated memory architecture.
We then introduce memory-disaggregated KV stores, focusing on their performance issues incurred by optimistic synchronization.
Finally, we introduce existing lock-based pessimistic synchronization approaches on DM and conduct preliminary experiments to show the potential performance boost of adopting them on memory-disaggregated KV stores.

\begin{figure*}[t]
    \centering
    \begin{minipage}[t]{0.405\columnwidth}
        \centering
        \includegraphics[width=\columnwidth]{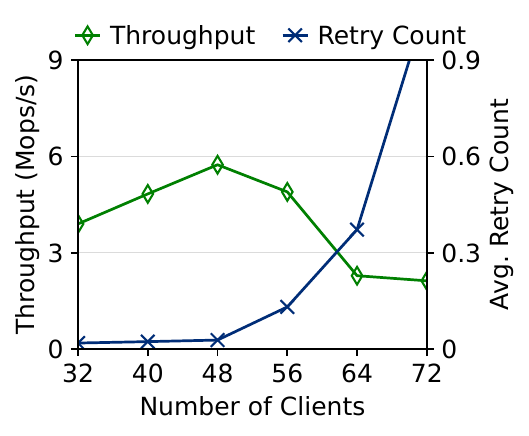}
        \caption{The throughput and retry count of the pointer array with optimistic synchronization under a highly-contented write-intensive workload.}
        \label{fig:fg01}
    \end{minipage}
    \hspace{3mm}
    \begin{minipage}[t]{0.362\columnwidth}
        \centering
        \includegraphics[width=\columnwidth]{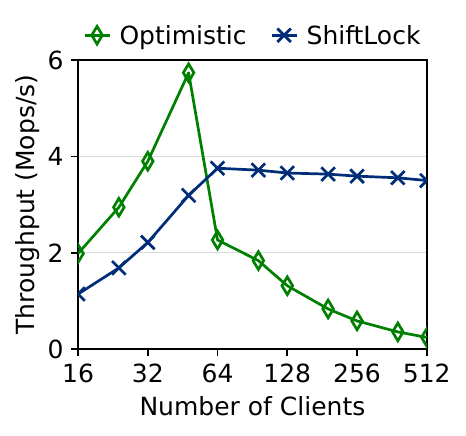}
        \caption{The throughput of the pointer array with optimistic and pessimistic synchronizations as a function of the number of clients.}
        \label{fig:fg02}
            
    \end{minipage}
    \hspace{3mm}
        \begin{minipage}[t]{0.362\columnwidth}
        \centering
        \includegraphics[width=\columnwidth]{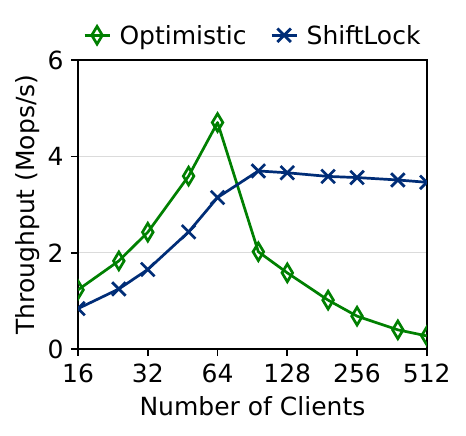}
        \caption{The throughput of RACE with optimistic and pessimistic synchronizations as a function of the number of clients.}
        \label{fig:fg03}
    \end{minipage}
    \hspace{3mm}
      \begin{minipage}[t]{0.375\columnwidth}
        \centering
        \includegraphics[width=\columnwidth]{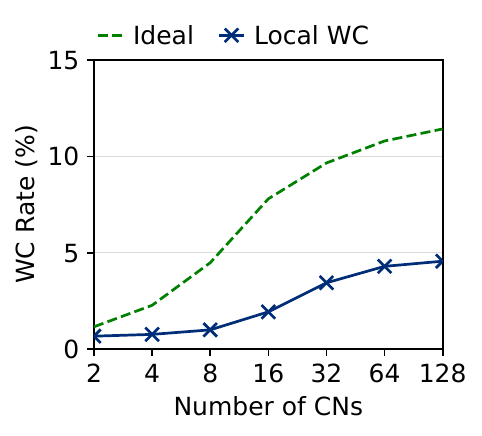}
        \caption{The actual WC rate of local WC and the theoretical upper bound as a function of the number of CNs.}
        \label{fig:fg04}
    \end{minipage}
    \hspace{3mm}
    \begin{minipage}[t]{0.375\columnwidth}
        \centering
        \includegraphics[width=\columnwidth]{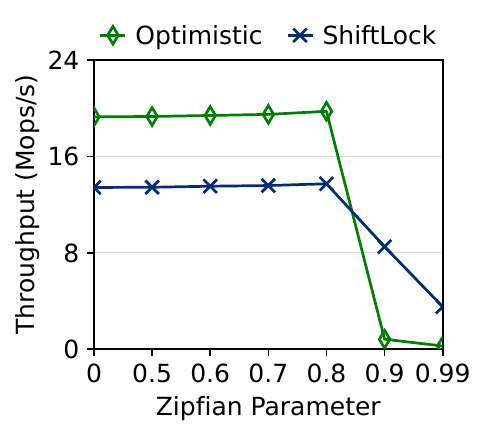}
        \caption{The throughput of optimistic and pessimistic synchronizations as a function of the skew factor (larger values mean more skew).}
        \label{fig:fg05}
    \end{minipage}
\end{figure*}

\subsection{Disaggregated Memory}
Disaggregated memory was proposed to attack the resource inefficiency issue due to the coupled allocation of CPU and memory in monolithic servers~\cite{guo2022clio,lee2021mind,shan2018legoos, Ruan2020aifm,wang2021concordia,nitu2018zbl,maruf2023dm,amaro2023dm}.
DM decouples computing and memory resources from monolithic servers into independent compute and memory pools.
Compute nodes (CNs) in the compute pool have abundant CPUs but only a limited amount of local memory to serve as the runtime cache.
Memory nodes (MNs) in the memory pool host sufficient memory capacity but only a few weak CPU cores to execute lightweight management computation, \ie network connection management and memory allocation management.
The compute and memory pools are connected with high-performance data center interconnect techniques, \eg RDMA~\cite{infiniband} and CXL~\cite{cxl}.
CNs can thus directly access and modify in-memory data on MNs, bypassing their weak CPUs.
In this paper, without loss of generality, we assume CNs access MNs with one-sided RDMA operations, \ie RDMA\_READ, RDMA\_WRITE, atomic compare and swap (RDMA\_CAS), and atomic fetch and add (RDMA\_FAA).
Other interconnect techniques that provide the same interfaces are compatible with our design.

\subsection{Memory-Disaggregated Key-Value Stores}\label{sec:background-dm-storage}
Many works port in-memory KV stores to DM to achieve better resource efficiency and elasticity~\cite{lee2023dinomo,shen2023fusee,shen2023ditto,Aguilera2023ubft}.
Existing memory-disaggregated KV stores are designed for simplicity and speed in accessing individual data with their keys~\cite{zuo2021race,luo2023smart,wang2022sherman,luo2024chime}.
They organize stored objects as key-value (KV) pairs and provide clients with standard KV interfaces, \ie \texttt{SEARCH(K)}, \texttt{INSERT(K, V)}, \texttt{UPDATE(K, V)}, and \texttt{DELETE(K)}~\cite{zuo2021race,shen2023fusee,li2023rolex}.
Complex operations, \eg transactions, are not usually supported.
Meanwhile, clients are responsible for handling invalid operations, \ie any attempt to \texttt{INSERT} an existing key, or to \texttt{SEARCH}, \texttt{UPDATE}, or \texttt{DELETE} a non-existent key will not be executed, and the client will receive a return value indicating that the operation is invalid~\cite{memcached,redis}.
A global index data structure is maintained in the memory pool to keep track of the mapping between the keys and values.

Memory-disaggregated KV stores are featured for their highly skewed and concurrent workloads~\cite{zuo2021race,luo2023smart,wang2022sherman,luo2024chime}.
Efficiently synchronizing clients in the compute pool to concurrently access data in the memory pool is critical to achieving high performance.
Existing approaches typically adopt an optimistic synchronization scheme.
They store pointers of KV pairs in their global index and synchronize concurrent data accesses in a lock-free manner~\cite{zuo2021race,min2024sephash,chen2020clevel}.
When performing \texttt{SEARCH} operations, clients on CNs first search for the target data pointer in the index with a series of RDMA operations.
Then the KV pair is fetched according to the pointer with RDMA\_READ.
When performing \texttt{INSERT}, \texttt{DELETE}, and \texttt{UPDATE} operations (IDU), they first write the modified KV pair to a new location in the memory pool and then atomically modify the data pointer in the global index with RDMA\_CAS.
The atomicity of RDMA\_CAS ensures that only one client can successfully modify the data pointer, and the values of pointers are always consistent. 
Consequently, \texttt{SEARCH} operations can always get the correct pointer and execute concurrently with IDU operations.

Compared with pessimistic synchronization, optimistic synchronization eliminates two additional remote memory accesses on acquiring and releasing locks in the memory pool. 
The operation latency can be reduced due to the reduced number of network round-trip times (RTTs) on the critical path.
Overall system throughput can also be improved in low contention workloads due to the lower per-operation overhead of IDU operations.

However, under highly contented workloads, optimistic synchronization becomes a significant performance bottleneck.
Specifically, when multiple clients concurrently IDU the same KV pair, only one thread can succeed due to the atomicity of RDMA\_CAS.
Other threads have to iteratively retry their operations until they successfully modify the data pointer.
Suppose, in the worst-case, when $n$ threads concurrently IDU the same pair and are acting in perfect synchrony, there will be $n$ rounds of retry operations since only one thread can succeed in each round.
However, during the process, $\frac{n(n-1)}{2}$ retry operations will be generated, causing $O(n^2)$ redundant operations.
These retries quickly saturate the limited IOPS and bandwidth of the memory pool.
System throughput is thus compromised since other operations are blocked by these redundant retries.
This is a severe issue for practical memory-disaggregated KV stores since real-world workloads typically follow Zipfian distributions with a high skewness~\cite{yang2020large,atikoglu2012workload}. 
The high skewness leads to severe contention during concurrent execution.

We integrate optimistic synchronization into a pointer array to illustrate its issues in high-contention scenarios.
Specifically, the pointer array is composed of 60 million pointers.
Each pointer stores the address of a unique KV pair in the memory pool.
Clients in the compute pool perform \texttt{SEARCH} and \texttt{UPDATE} operations on the pointer array with optimistic synchronization.
We evaluate the throughput and number of retried operations of the pointer array under a write-intensive workload with 50\% \texttt{SEARCH} and 50\% \texttt{UPDATE}.
Requests are generated following a Zipfian distribution with $\theta=0.99$ to reflect real-world workloads with skewed data access patterns.
As shown in Figure~\ref{fig:fg01}, with less than 48 clients, the throughput of the pointer array scales linearly, and the number of retried operations stays low.
This is because the pointer array is bottlenecked by the computing power of clients.
With more than 48 clients, the throughput drops by up to $2.7\times$ as the number of clients increases.
This can be explained by the dramatic increase in the number of retried operations generated when handling conflicting \texttt{UPDATE} operations.
These retry operations quickly saturate the RDMA network interface cards (RNICs) in the memory pool, preventing other operations from being normally executed.

\subsection{Pessimistic Synchronization on DM}\label{sec:backgound-pessimistic}
The spinlock is the most widely adopted pessimistic synchronization approach on DM.
It is typically implemented by iteratively polling a remote memory address with atomic RDMA\_CAS~\cite{luo2023smart,wang2022sherman,luo2024chime}.
The key problem with spinlocks on DM lies in their expensive polling overhead incurred on the RNICs of the memory pool.

ShiftLock~\cite{gao2025shiftlock} is the state-of-the-art lock mechanism on DM.
ShiftLock ports the \textbf{M}ellor-\textbf{C}rummey-\textbf{S}cott (MCS) lock~\cite{mellor1991mcs,michael2013book}, which is originally designed for NUMA systems, to DM to relieve the network bottleneck of the memory pool.
Specifically, ShiftLock maintains each lock as a linked list.
All clients waiting for the same lock are queued in the same linked list, indicating the order of getting locks.
For each lock, a lock entry is maintained in the memory pool to record the client ID at the tail of the list.
When a client acquires a lock, it appends itself to the list by first executing an atomic get-and-set operation on the corresponding lock entry.
The get-and-set operation is implemented with masked RDMA\_CAS by setting the $compare\_mask$ to 0 to bypass the comparison in a normal RDMA\_CAS operation~\cite{masked-cas,gao2025shiftlock}.
With the previous tail in the return value of the get-and-set operation, the client finishes the lock acquisition by notifying the previous tail about the insertion of a new client.
When a client releases a lock, it transfers the ownership of the lock to the next client by informing the following client in the list with a network request.
The polling overhead on memory-side RNICs is thus shifted to client-side RNICs, which significantly improves system performance.

\textbf{Motivation.}
In this paper, we propose to adopt lock-based pessimistic synchronization to enhance the performance of memory-disaggregated KV stores.
To validate the potential of using pessimistic synchronization to address the limitations of optimistic synchronization under highly contended workloads, we incorporate the MCS lock of ShiftLock into KV stores with different index structures, \eg the pointer array and the RACE hash~\cite{zuo2021race}. 
Figure~\ref{fig:fg02} shows the throughput of the pointer array with optimistic and pessimistic synchronization under the same workload as in Section~\ref{sec:background-dm-storage}.
With optimistic synchronization, the throughput of the pointer array significantly drops after reaching the peak at 48 clients.
However, with pessimistic synchronization, \ie ShiftLock~\cite{gao2025shiftlock}, the performance remains stable.
As shown in Figure~\ref{fig:fg03}, a similar trend is observed on RACE hashing~\cite{zuo2021race}, indicating that pessimistic synchronization also works on real-world applications.

\section{Challenges}\label{sec:analysis}
The performance of memory-disaggregated KV stores after adopting pessimistic synchronization is still suboptimal due to two challenges: 1) severe cross-node redundant data modifications and 2) high lock maintenance overhead.
In this section, we introduce these two challenges in detail with thorough experimental analyses.

\subsection{Inter-Node Redundant Data Modifications}
Even though pessimistic synchronization effectively reduces redundant I/Os during data synchronization, substantial I/O redundancy still exists during concurrent IDU operations.
Specifically, when multiple clients concurrently execute IDU operations on the same KV pair, all of them first write the modified KV pair to a new location in the memory pool with RDMA\_WRITE.
They then update the same pointer in the global index data structure with RDMA\_CAS.
Since the data pointer is protected by a lock, the update of the pointer happens sequentially.
Most of the concurrent IDU operations in the process are wasted since they will be quickly overwritten by subsequent IDU operations to the same KV pair, resulting in wasted RDMA\_WRITE and RDMA\_CAS operations.
Such a problem is particularly acute in production environments since real-world workloads are usually skewed~\cite{cooper2010benchmarking}.

Existing approaches adopt a local write-combining (WC) technique to reduce redundant I/Os incurred by concurrent IDU operations~\cite{luo2023smart,luo2024chime}.
The key idea is to consolidate concurrent IDU operations to the same KV pair in a small time window into one IDU operation.
To detect redundant IDU operations during the time window, existing approaches associate a lock with each KV pair locally on each CN.
The client that successfully acquires the local lock becomes the combiner and executes the IDU operation to the memory pool.
Concurrent clients that issue IDU operations to the same KV pair are all blocked by the local lock and combined by the combiner.
They directly return the combiner's result after successfully acquiring the lock since their operations are already executed.
To ensure the correctness of the combined KV pair, existing approaches adopt a last-writer-wins conflict resolution scheme~\cite{Lynch1997lww,luo2023smart}.
A WC buffer is maintained locally on each CN for each KV pair.
All blocked clients write their modified KV pair to the WC buffer by overwriting the KV pair written by previous clients.
Consequently, the WC buffer always contains the KV pair written by the last writer.
The combiner finally writes the KV pair to the memory pool, which indicates the value written by the last writer.

However, local WC can only handle redundant I/Os of IDU operations issued by clients within the same CNs.
There are still extensive amounts of redundant I/Os generated by clients across CNs.
To illustrate this issue, we integrate local WC to the pointer array described in Section~\ref{sec:background-dm-storage} and evaluate its WC rate with increasing numbers of CNs.
Since we only have access to a limited number of physical servers, we simulate a large number of virtual CNs by partitioning our servers.
Specifically, each virtual CN consists of 4 CPU cores, and we run a client on each core, thus allowing a maximum of 4 concurrent clients on our CNs.
The WC rate is calculated as $\frac{\# combined\_requests}{\# total\_requests}$.
We also evaluate the redundant request rate to show an upper bound for the WC rate, \ie the ideal WC rate.
The redundant request rate is calculated as the $\frac{\# combined\_requests \; + \; \# blocked\_requests}{\# total\_requests}$, where $\#blocked\_requests$ is the number of requests blocked by lock-based pessimistic synchronization.
As shown in Figure~\ref{fig:fg04}, both the redundant request rate and the local WC rate increase with the number of CNs.
Approximately 7\% of redundant requests are not combined. These requests predominantly access hot keys and exhibit high latency. Since requests for hot keys must be serialized through lock acquisition, their throughput is constrained by latency. Consequently, this 7\% of requests significantly degrades both system throughput and tail latency.

\subsection{High Lock Maintenance Overhead}
The second challenge lies in the high lock maintenance overhead.
When leveraging lock-based pessimistic synchronization, clients have to acquire and release a lock before and after performing IDU operations.
Both acquiring and releasing locks are remote memory accesses since locks are maintained in the memory pool to synchronize all clients in the compute pool.
This incurs two additional RTTs on the critical path of each IDU operation.
The system throughput is thus compromised in low contention workloads due to the higher per-operation latency and I/O overhead.

We verify the insufficiency of naively adopting pessimistic synchronization by evaluating the performance of the pointer array with optimistic and pessimistic synchronization under various contention scenarios.
The contention can be affected by both the number of concurrent clients and the skewness of the workload.

First, under the same number of clients, the higher the skewness, the higher the contention.
Figure~\ref{fig:fg05} shows the performance of the pessimistic and optimistic pointer arrays with 512 clients under the write-intensive workload with 50\% \texttt{SEARCH} and 50\% \texttt{UPDATE}.
We use different $\theta$s of the Zipfian distribution to control the skewness of the workload, \ie a larger $\theta$ indicates a more skewed workload.
The throughput of pessimistic synchronization is only about $70\%$ that of the optimistic approach with $\theta \le 0.8$ due to the higher per-operation I/O overhead.
However, under more skewed workloads, the throughput of pessimistic synchronization is up to $14\times$ better since there are fewer redundant I/Os generated by the retry operations of optimistic synchronization.

Moreover, under the same skewness, the more clients, the higher the contention.
Our previous results in Figure~\ref{fig:fg02} show a similar trend, \ie optimistic synchronization performs better with a small number of clients, while pessimistic synchronization performs better when the number of clients is larger.


\section{The \DMCS Design}

\begin{figure}[!t]
    \centering
    \includegraphics[width=0.95\linewidth]{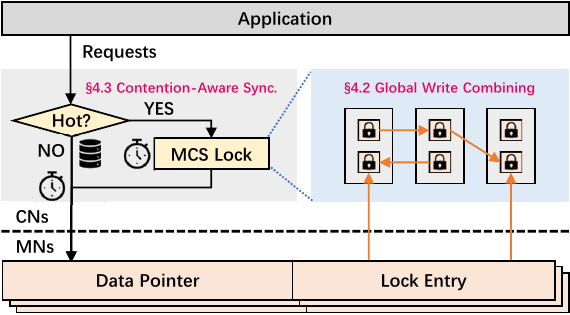}
    \caption{The overview of \DMCS.}\label{fig:overview}
\end{figure}

\subsection{\DMCS Overview}\label{sec:overview}
We propose \DMCS, a compute-side I/O optimization framework that leverages pessimistic synchronization to boost existing memory-disaggregated KV stores that conduct \texttt{INSERT}, \texttt{DELETE}, \texttt{UPDATE}, and \texttt{SEARCH} on data pointers as defined in Section~\ref{sec:background-dm-storage}.
Figure~\ref{fig:overview} shows the overview of \DMCS.
\DMCS adopts the distributed MCS lock in ShiftLock~\cite{gao2025shiftlock} to achieve pessimistic synchronization.
We disable the reader-writer lock optimization of ShiftLock since it adds additional overhead on single-key accesses.
To reduce the number of redundant \texttt{UPDATE} operations (\textbf{Challenge 1}), we propose a global write-combining (WC) technique that leverages MCS locks to detect and consolidate concurrent and redundant \texttt{UPDATE} operations across CNs.
To improve pessimistic synchronization in low-contention scenarios (\textbf{Challenge 2}), we propose a contention-aware synchronization scheme that adaptively switches between pessimistic and optimistic synchronization schemes according to the contention level of individual keys.
We introduce these two techniques in \S~\ref{sec:global-write-combining-on-dm} and \S~\ref{sec:contention-aware-synchronization}, respectively.

\subsection{Global Write Combining}\label{sec:global-write-combining-on-dm}
Global WC is proposed to reduce cross-node redundant I/Os by aggregating IDU requests from multiple CNs.
Compared with local WC, the key challenge of global WC lies in efficiently detecting and correctly combining redundant IDU operations, \ie concurrent IDU operations that modify the same KV pair.
One straightforward approach is to adopt a centralized coordination server that processes all IDU operations and combines redundant ones. 
However, the centralized server can become a severe performance bottleneck when the number of clients grows, making the system hard to scale to a large number of clients.

The key opportunity to address this issue lies in the MCS lock mechanism. 
For each KV pair, the MCS lock maintains a distributed wait queue across CNs, obviating the need for an additional centralized server to detect concurrent and redundant requests.
Moreover, since the order of clients in the wait queue indicates the order of performing IDU operations, the operations can be safely combined with a last-writer-wins conflict resolution scheme~\cite{shen2023fusee,Lynch1997lww}.

Based on this observation, we propose global WC over the MCS lock.
In the rest of this section, we first introduce how \DMCS achieves \texttt{UPDATE} operations in Section~\ref{sec:workflow-global-wc}.
\texttt{INSERT} and \texttt{DELETE} operations are separately introduced in Section~\ref{sec:corner-case} since they are handled differently to guarantee correctness.

\begin{figure}[!t]
    \centering
    \includegraphics[width=0.84\linewidth]{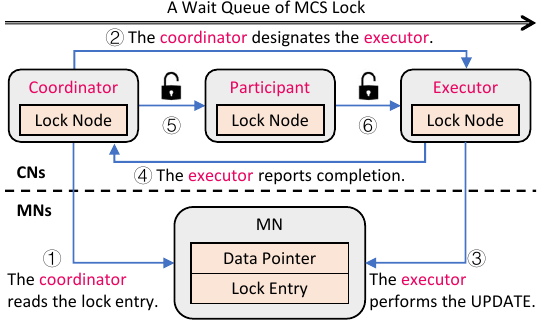}
    \caption{The workflow of global WC.}\label{fig:gwc}
\end{figure}

\subsubsection{Combining \texttt{UPDATE} operations}\label{sec:workflow-global-wc}
Figure~\ref{fig:gwc} presents the workflow of global WC over \texttt{UPDATE} operations.
Among CNs, clients concurrently executing \texttt{UPDATE} operations on the same KV pair are organized into the same wait queue with a lock node.
Inside the queue, \DMCS distinguishes three types of clients, \ie coordinators, participants, and executors.
Coordinators are clients who initiate a global WC.
Executors are clients that actually execute the operations with remote memory accesses.
Finally, participants are other clients who are combined by the executor.
On MNs, \DMCS associates each data pointer with a lock entry to synchronize concurrent modifications to the corresponding KV pair.

\begin{figure}[!t]
    \centering
    \includegraphics[width=\linewidth]{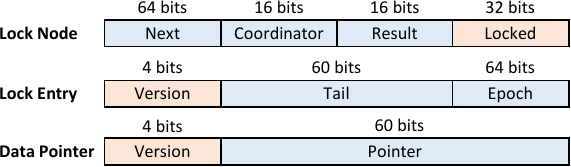}
    \caption{The structures of the lock node, lock entry, and data pointer.}\label{fig:lockstructure}
\end{figure}

Figure~\ref{fig:lockstructure} shows the detailed structures of the lock node, lock entry, and data pointer.
On CNs, each lock node contains a 64-bit \texttt{Next} field to indicate the next client in the wait queue, a 16-bit \texttt{Coordinator} field to record the coordinator of global WC, a 16-bit \texttt{Result} field to propagate the result of the combined operation, and a 32-bit \texttt{Locked} field to handover the ownership of the lock.
Data pointers on MNs contain 64 bits with a 60-bit \texttt{Pointer} and a 4-bit \texttt{Version} to handle \texttt{DELETE} operations, which will be discussed later (\S~\ref{sec:corner-case}).
Lock entries serve as gateways for individual MCS locks.
They contain a 60-bit \texttt{Tail} to store the client ID at the tail of the current wait queue and a 64-bit \texttt{Epoch} for fault tolerance.
A 4-bit \texttt{Version} is also maintained to deal with \texttt{DELETE} operations.

Global WC starts when a client successfully acquires the lock.
The client then checks if there are other \texttt{UPDATE} operations that can be combined before modifying the data pointer.
It achieves this by examining if the  \texttt{Next} field in its local lock node is modified by subsequent clients.
If the \texttt{Next} field remains empty, the client becomes an executor and normally finishes the \texttt{UPDATE}.
Otherwise, the client becomes a coordinator and starts combining subsequent operations.
The coordinator first identifies the executor, \ie the client on the tail of the wait queue, by reading the lock entry on the MN (Step \textcircled{\footnotesize 1}).
The coordinator then notifies the executor to start combining operations by transferring ownership of the lock to the executor and sending it the coordinator's ID through modifying its \texttt{Coordinator} field (Step \textcircled{\footnotesize 2}).
After getting the ownership of the lock, the executor finishes the \texttt{UPDATE} operation by first writing the updated KV pair to a newly allocated location and then modifying the data pointer (Step \textcircled{\footnotesize 3}).
The executor then notifies the coordinator of the result of the combined \texttt{UPDATE} operations and transfers the ownership of the lock back (Step \textcircled{\footnotesize 4}).
The coordinator then skips its \texttt{UPDATE} operation, returns the result to upper-level applications, and transfers the ownership of the lock and the result along the wait queue (Step \textcircled{\footnotesize 5}).
The result of the combined \texttt{UPDATE} is conveyed by modifying the \texttt{Result} field in the lock node.
Meanwhile, the \texttt{Locked} field is set to a special value \texttt{0x3} to indicate that the client's \texttt{UPDATE} is combined by the executor.
All subsequent participants also skip their \texttt{UPDATE} operations and propagate the WC result through successive lock transfers when finding the \texttt{Locked} field is set to \texttt{0x3} (Step \textcircled{\footnotesize 6}).

Compared with only adopting the standard MCS lock, global WC introduces an additional remote memory access overhead to identify the executor (Step \textcircled{\footnotesize 1}) and two cross-CN communications (Steps \textcircled{\footnotesize 2} and \textcircled{\footnotesize 4}).
However, it ensures that only a single \texttt{UPDATE} is actually executed on MN per combined batch, regardless of the number of requests in the batch.
This design fundamentally eliminates cross-node redundant I/Os, especially I/Os on MNs.
System performance can thus be improved under highly concurrent workloads with skewed access patterns.

\subsubsection{Combining \texttt{INSERT} and \texttt{DELETE} operations}\label{sec:corner-case}
Handling \texttt{INSERT} and \texttt{DELETE} operations is difficult for global WC due to the mismatch between the targets of locks and those of global WC.
Specifically, all concurrent operations modifying the same data pointer will be organized in the same wait queue, while only operations on the same KV pair can be combined.
\texttt{INSERT} and \texttt{DELETE} operations can alter the key associated with the same data pointer, compromising the correctness of global WC.

\DMCS ensures that operations in the same wait queue modify the same key by handling \texttt{INSERT} and \texttt{DELETE} operations differently.
For \texttt{INSERT} operations, clients directly write the new KV pair to the memory pool and modify the empty pointer with RDMA\_CAS without entering the MCS wait queue.
For \texttt{DELETE} operations, \DMCS ensures they are always the last operations on their corresponding wait queues.
This ensures that all operations in the queue target the same KV pair and will be combined by the last \texttt{DELETE}.

\DMCS leverages the version numbers in lock entries and data pointers to deal with \texttt{DELETE} operations.
When performing a \texttt{DELETE} operation, a client first increments the version number in the lock entry when acquiring locks.
The version number in the data pointer is incremented after the \texttt{DELETE} finishes.
When performing \texttt{UPDATE} operations, clients first read the data pointer to get the version number and the address of the old KV pair. 
Then the client acquires the lock with the version number in the data pointer.
The lock acquisition operation will be rejected when there is a mismatch between the two version numbers, indicating the target KV pair is already deleted.
\texttt{UPDATE} operations initiated in between the start and finish of the \texttt{DELETE} operations are thus rejected from the wait queue, ensuring the correctness of global WC.

\subsection{Contention-Aware Synchronization}\label{sec:contention-aware-synchronization}

As we discussed in Section~\ref{sec:analysis}, optimistic and pessimistic synchronizations exhibit complementary performance characteristics under workloads with different levels of contentions.
This motivates us to design contention-aware synchronization to adapt synchronization schemes dynamically, matching the contention level of workloads.

A straightforward approach to achieve this is to adopt a centralized manager process.
The manager samples requests from CNs to MNs to monitor the level of contention of the current workload.
When the observed contention level changes significantly, it pauses new requests, switches the concurrency control protocol, and resumes operation.
However, this centralized and coarse-grained approach poses several inherent limitations.
First, there is a fundamental trade-off between the monitoring overhead and estimation accuracy, \ie higher sampling frequency improves the accuracy of load estimation but introduces additional system overhead.
Second, switching between synchronization schemes requires pausing all clients, which impacts service stability and increases tail latency.

We propose to \textit{dynamically decide synchronization modes with a decentralized and fine-grained arbitration mechanism} to avoid maintaining an additional centralized manager process.
The key idea is similar to credit-based throttling~\cite{monga2021credit}.
Clients monitor the performance statistics of executing each IDU operation and assign each data pointer credits.
A higher credit indicates a higher contention level.
When executing IDU operations, clients adaptively decide to adopt optimistic or pessimistic synchronization according to the credit of the data pointer.
Such a decentralized approach can effectively perceive data access patterns and adapt to workload change without the overhead induced in a centralized method.

\begin{algorithm}[t]
\SetKwFunction{MyFunction}{update}
\SetKwFunction{FuncOpti}{optimisticUpdate}
\SetKwFunction{FuncPess}{pessimisticUpdate}
\SetKwProg{Fn}{Function}{:}{\KwRet} 
\newcommand{\mycommentstyle}{\color{mygraygreen}\itshape}
\SetCommentSty{mycommentstyle}

\SetKwInOut{GV}{Global Variables}
\SetKwInOut{myInput}{Algorithm Inputs}
\GV{

    \textcolor{mygraygreen}{\textit{/* The following two variables are maintained on each CN. */ }}

    $credit$: A hash map recording the credit count.

    $retryRecord$: A hash map tracking the number of retries.
}

\myInput{

  $ptrAddr$: The address of the target data pointer.

  $KV$: The KV pair to be updated.

  $oldPtr$: The value of the data pointer.
}

\Fn{\MyFunction{$ptrAddr,KV,oldPtr$}}{
    \eIf{ $credit[ptrAddr] > 0$ }{ 
        $credit[ptrAddr]\mathrel{-}=$ 1 \;
        \FuncPess{$ptrAddr,KV,oldPtr$}
    }{
        \FuncOpti{$ptrAddr,KV,oldPtr$}
    }
}
\Fn{\FuncPess{$ptrAddr,KV,oldPtr$}}{
    $WCResult \gets$ $ptrAddr.$tryLockAndWC()\;
    \If { the client becomes the executor of global WC } {
     \tcc{Allocate a new memory block, write the KV pair to it with RDMA\_WRITE, and return the address of the newly written KV.}
        $newPtr \gets$ allocateAndWriteKV($KV$)\;
    \tcc{Update data pointer with RDMA\_CAS, record and return the number of retries.}
        $nRetry \gets$ CASPointer($ptrAddr,oldPtr,newPtr$)\;
        $ptrAddr.$unlock()\;
    }
    \tcc{In all other cases, $ptrAddr$ will NOT lock.}
    \eIf{ $WCBatchSize > 1$}{
        $credit[ptrAddr] \mathrel{+}= 2$\;
    }{
        $credit[ptrAddr] \mathrel{/\!}= AIMD\_FACTOR$\;
    }
}
\Fn{\FuncOpti{$ptrAddr,KV,oldPtr$}}{
    $newPtr \gets$ allocateAndWriteKV($KV$)\;
    $nRetry \gets$ CASPointer($ptrAddr,oldPtr,newPtr$)\;
    \If{ $nRetry >= HOTNESS\_THRESHOLD$ \&\& $retryRecord[ptrAddr] >= HOTNESS\_THRESHOLD$}{
        $credit[ptrAddr] \mathrel{+}= $ INITTIAL\_CREDIT\;
    }
    $retryRecord[ptrAddr] \gets nRetry$\;
}

\caption{Contention-Aware Synchronization} \label{alg:contention-aware-sync}
\end{algorithm}

Algorithm~\ref{alg:contention-aware-sync} shows the process of contention-aware synchronization, which involves changes only to the final step of the update operation, \ie the modification of data pointers.
Each CN maintains two hash maps to track the contention level of workloads, \ie a \textit{credit} map to record credits of each data pointer and a \textit{retryRecord} map to record retry counts associated with each data pointer.

When updating a data pointer, a client first decides the synchronization mode by checking whether the target data pointer possesses any credits.
If so, it consumes one credit and proceeds with pessimistic synchronization (Lines 3-4).
In the pessimistic mode, global WC is adopted to reduce redundant data modifications.
If the client becomes the executor of global WC, it needs to first write the KV and then update the pointer with an RDMA\_CAS operation (Lines 9-12). In all other cases, the client will not hold the lock and is blocked until the combined request has been processed.
Finally, the client dynamically adjusts the number of credits based on congestion assessment, which is determined by the batch size of global WC (Lines 13-16).

Otherwise, if the target data pointer does not contain any credit, it executes the operation with optimistic synchronization since the target key is infrequently accessed (Lines 18-19). Similarly, the client evaluates congestion by comparing the number of CAS retries between the current and previous attempts, and adjusts credits accordingly (Lines 20-22).

To promptly detect contention level variations, we adopt the additive increase and multiplicative decrease scheme (AIMD)~\cite{chiu1989aimd} to update credits. Specifically, we increase the credit after a successful global WC, while dividing the credit by an \texttt{AIMD\_FACTOR} when no concurrent requests are detected for global WC (Lines 13-16). The \texttt{AIMD\_FACTOR} is empirically set to 2, which is consistent with the original algorithm~\cite{chiu1989aimd}. For optimistic synchronization, if two consecutive requests require two or more retries (\ie the \texttt{HOTNESS\_THRESHOLD} is set to 2), we increase the credit by \texttt{INITIAL\_CREDIT} (Lines 20-21), which is empirically set to 36.

The contention-aware synchronization scheme enables fine-grained and seamless transitions between synchronization modes for individual KV pairs.
For hot KV pairs, pessimistic synchronization with global WC eliminates redundant IDU operations by batching modifications and preventing unnecessary retries. 
For cold KV pairs, optimistic synchronization avoids the expensive lock maintenance overhead and reduces operation latency.
Moreover, contention-aware synchronization can also improve the efficiency of global WC.
The performance gain of global WC increases when combining a large batch of requests, since the fixed overhead of global WC can be amortized.
Contention-aware synchronization boosts the performance of global WC by combining only IDU operations for hot KV pairs, which usually results in larger batches of operations being combined.

\begin{figure}[t]
   \centering
    \includegraphics[width=0.97\linewidth]{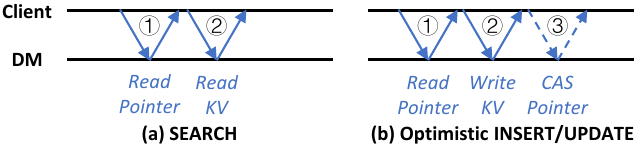}
    \caption{The workflows of \texttt{SEARCH}, \texttt{INSERT}, and \texttt{UPDATE} operations under the optimistic mode.}\label{fig:atgd1}
\end{figure}

\begin{figure*}[t]
   \centering
    \includegraphics[width=\textwidth]{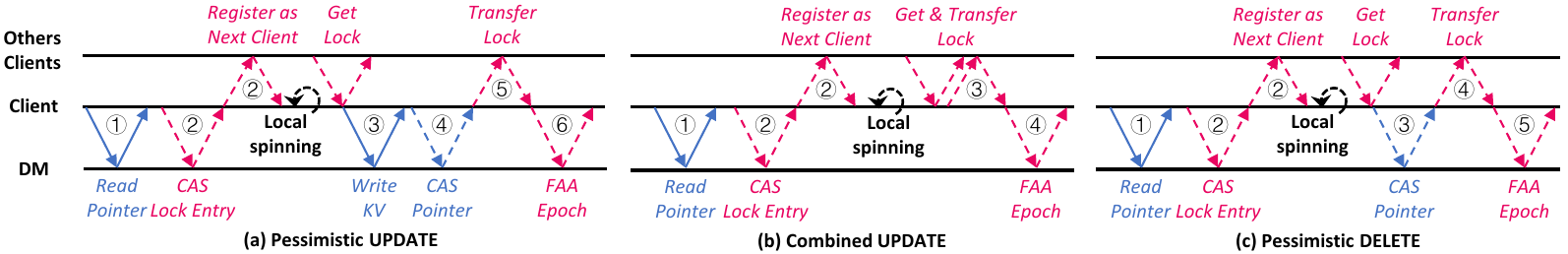}
    \caption{The workflow of the \texttt{UPDATE} and \texttt{DELETE} operations under the pessimistic mode. The solid lines indicate RDMA\_READ and RDMA\_WRITE, and the dashed lines indicate atomic RDMA\_CAS and RDMA\_FAA.}
    \label{fig:atgd2}
\end{figure*}

        
  

\subsection{Put It All Together}
\DMCS is compatible to all memory-disaggregated systems with optimistic out-of-place modification in their index data structures~\cite{luo2023smart,li2023rolex,zuo2021race,min2024sephash}.
This section introduces the \texttt{SEARCH}, \texttt{INSERT}, \texttt{UPDATE}, and \texttt{DELETE} operations of \DMCS in detail. 

As shown in Figures~\ref{fig:atgd1} and \ref{fig:atgd2}, each operation begins by querying the index structure to locate the target data pointer (\textcircled{\footnotesize 1}). 
The final RDMA\_FAA for each operation in Figure~\ref{fig:atgd2} is used for fault tolerance, which will be described in Section~\ref{sec:discussion}.

\textbf{Search.} 
\DMCS enables lock-free \texttt{SEARCH} operations as shown in Figure~\ref{fig:atgd1}a.
The client first reads the data pointer via an RDMA\_READ (\textcircled{\footnotesize 1}).
It then performs an RDMA\_READ on the memory location referenced by the pointer, where the KV pair is stored (\textcircled{\footnotesize 2}).

\textbf{Insert.} 
\DMCS always conducts \texttt{INSERT} operations with optimistic synchronization as shown in Figure~\ref{fig:atgd1}b.
The client first writes the new KV data to a newly allocated memory region on the MN via an RDMA\_WRITE (\textcircled{\footnotesize 2}).
It then uses RDMA\_CAS to atomically modify the data pointer to point to the new data (\textcircled{\footnotesize 3}).

\textbf{Update.}
\DMCS's optimizations primarily focus on the \texttt{UPDATE} operation.
The client first selects the suitable synchronization mode based on the contention-aware synchronization scheme introduced in Section~\ref{sec:contention-aware-synchronization}.
In the optimistic mode, as shown in Figure~\ref{fig:atgd1}b, the client writes the new KV data to a new memory region on the MN (\textcircled{\footnotesize 2}) and then executes an RDMA\_CAS operation to atomically swap the old data pointer with the new memory address (\textcircled{\footnotesize 3}).
If the RDMA\_CAS fails, the client retries the update operation.

In the pessimistic mode, as shown in Figure~\ref{fig:atgd2}a and \ref{fig:atgd2}b, the client acquires the MCS lock associated with the data pointer (\textcircled{\footnotesize 2}) and tries to participate in the global WC process introduced in Section~\ref{sec:global-write-combining-on-dm}.
When participating in WC as the coordinator or participant, as shown in Figure~\ref{fig:atgd2}b, the client directly returns the combined result upon lock release (\textcircled{\footnotesize 3}).
Otherwise, as shown in Figure~\ref{fig:atgd2}a, after acquiring the lock, the client conducts the remote out-of-place update, \ie writes the new KV data (\textcircled{\footnotesize 3}) and performs an RDMA\_CAS to update the pointer (\textcircled{\footnotesize 4}).
After that, the client releases the lock (\textcircled{\footnotesize 5}). 
The client adjusts the number of credits associated with the data pointers according to local information, \eg the retry count, to enable the client to select the better synchronization mode during the next \texttt{UPDATE} operation.

\textbf{Delete.} 
\DMCS always executes \texttt{DELETE} operations with pessimistic synchronization, as shown in Figure~\ref{fig:atgd2}c.
The client first acquires the MCS lock of the data pointer and increments the version number in the lock entry via an RDMA\_CAS (\textcircled{\footnotesize 2}).
It then atomically sets the data pointer to null and increments the version number in the data pointer via an additional RDMA\_CAS (\textcircled{\footnotesize 3}).
Finally, the client releases the MCS lock (\textcircled{\footnotesize 4}).

\subsection{Correctness}

\subsubsection{The correctness of global WC}
There are three key aspects that affect the correctness of global WC.
First, all combined operations must target the same KV pair.
\DMCS ensures this by never combining \texttt{INSERT} operations and always ensuring \texttt{DELETE} operations are the executors.
Second, all combined operations must be concurrent operations in a small time window.
\DMCS ensures the correctness by guaranteeing the same time window as the local WC.
The time window of global WC is defined as the period for the first client, \ie the coordinator, to successfully acquire the lock and identify the last client in the wait queue, \ie the executor, which is the same as local WC~\cite{luo2023smart}.
All \texttt{UPDATE} operations issued by clients outside the time window will be queued behind the executor and combined by the next executor.
Finally, all operations should be combined with a correct conflict resolution scheme.
Global WC adopts a last-writer-wins~\cite{Lynch1997lww} conflict resolution similar to that of local WC.
This is guaranteed by always identifying executors as the last clients in the wait queue, \ie the last writer.
Operations are thus correctly combined by only writing the value of the last writer to the memory pool.

\subsubsection{The correctness of contention-aware synchronization}
The correctness of contention-aware synchronization can be argued by correctly synchronizing \texttt{INSERT}, \texttt{DELETE}, \texttt{UPDATE}, and \texttt{SEARCH} operations in two synchronization modes.
Specifically, since IDU operations always modify data pointers with atomic RDMA\_CAS, \texttt{SEARCH} operations can always safely read the correct pointer with RDMA\_READ and fetch the corresponding KV pair.
When executing IDU operations in both pessimistic and optimistic modes, \DMCS always uses RDMA\_CAS to modify the data pointer.
Consequently, IDU operations in different synchronization modes can also be correctly synchronized due to the atomicity of RDMA\_CAS.

\subsection{Discussions}\label{sec:discussion}

\textbf{\textit{Fault tolerance.}}
The key fault-tolerance issue of \DMCS lies in the reliance on the MCS lock~\cite{gao2025shiftlock}.
Deadlocks can happen on client failures since the wait queue is constructed by organizing clients into a distributed linked list.
\DMCS adopts the same design as ShiftLock~\cite{gao2025shiftlock} to achieve fault tolerance. Specifically, ShiftLock assumes that operations that hold locks have a maximum duration. Consequently, deadlocks can be detected and repaired whenever a client is waiting too long.
Similar to ShiftLock, \DMCS associates each lock with a 64-bit \texttt{Epoch}, as shown in Figure~\ref{fig:lockstructure}.
When releasing a lock, clients increment the lock's \texttt{Epoch} field by one via an RDMA\_FAA. 
Deadlock is detected if the \texttt{Epoch} field of the lock exhibits no increase over a full maximum duration while clients are waiting for acquisition.
The maximum duration is a configurable parameter, and we set the value to be the same as ShiftLock.
Clients then report the deadlock to the MN, and the MN handles the failure by resetting the lock.

\noindent
\textbf{\textit{Generality.}}
\DMCS is designed to boost memory-disaggregated KV stores with optimistic synchronization.
A large amount of existing memory-disaggregated KV stores can be optimized with \DMCS since optimistic synchronization is widely adopted in the literature~\cite{zuo2021race, luo2023smart, li2023rolex, luo2024chime, min2024sephash}.
However, the applicability of \DMCS is not limited to memory disaggregated KV stores.
All applications on DM that adopt optimistic synchronization with out-of-place data modification can benefit from \DMCS.

\noindent
\textbf{\textit{Fairness.}}
\DMCS has better fairness compared with only adopting optimistic synchronization.
Specifically, optimistic synchronization on DM relies on atomic RDMA\_CAS to resolve conflicts. 
However, commercial RNICs do not guarantee fairness for RDMA\_CAS operations, \ie a client can repeatedly fail to modify a data pointer.
\DMCS improves this by introducing MCS lock-based pessimistic synchronization, which enforces clients to perform pointer modification in a FIFO order.

\noindent
\textbf{\textit{Metadata overhead.}}
\DMCS introduces additional metadata overhead on both CNs and MNs.
On CNs, the metadata overhead is incurred by storing \texttt{credit} and \texttt{retryRecord} associated with data pointers. 
Specifically, for each KV pair, the metadata overhead is 8 bytes, \ie 4 bytes for \texttt{credit} and 4 bytes for \texttt{retryRecord}.
\DMCS reduces this overhead by recording \texttt{credit} and \texttt{retryRecord} only for hot KV pairs.
On MNs, the metadata overhead is incurred by storing lock entries associated with data pointers.
Specifically, for each key-value pair, the metadata overhead totals 24 bytes, which matches the overhead in ShiftLock~\cite{gao2025shiftlock}. This consists of 8 bytes for data pointers and 16 bytes for the lock structure.
This metadata overhead is considered acceptable. For reference, popular key-value stores such as Memcached~\cite{memcached} also require a minimum of 31 bytes of metadata per key-value pair~\cite{bin2013mem3c}.


\section{Evaluation}

\renewcommand{\arraystretch}{1.1}
\begin{table}
    \centering
    \caption{Workload specifications.}
    \small
    \begin{tabular}{c|c|c}
         \Xhline{1.5pt}
          \textbf{Workload} & \textbf{Write Ratio (\texttt{IDU})} & \textbf{Read Ratio (\texttt{SEARCH})} \\
         \hline
        \hline
          \textbf{write-intensive} & 50\%  & 50\% \\
          \textbf{read-intensive}  & 5\%   & 95\%   \\
          \textbf{write-only}      & 100\% & /       \\
         \Xhline{1.5pt}
    \end{tabular}
    \label{tab:workload}
\end{table}

\begin{figure*}[!t]
    \begin{minipage}[t]{0.47\textwidth}
        \centering
    \begin{tikzpicture}
        \node[anchor=south west, inner sep=0] (image) at (0,0) {
            \includegraphics[width=\columnwidth]{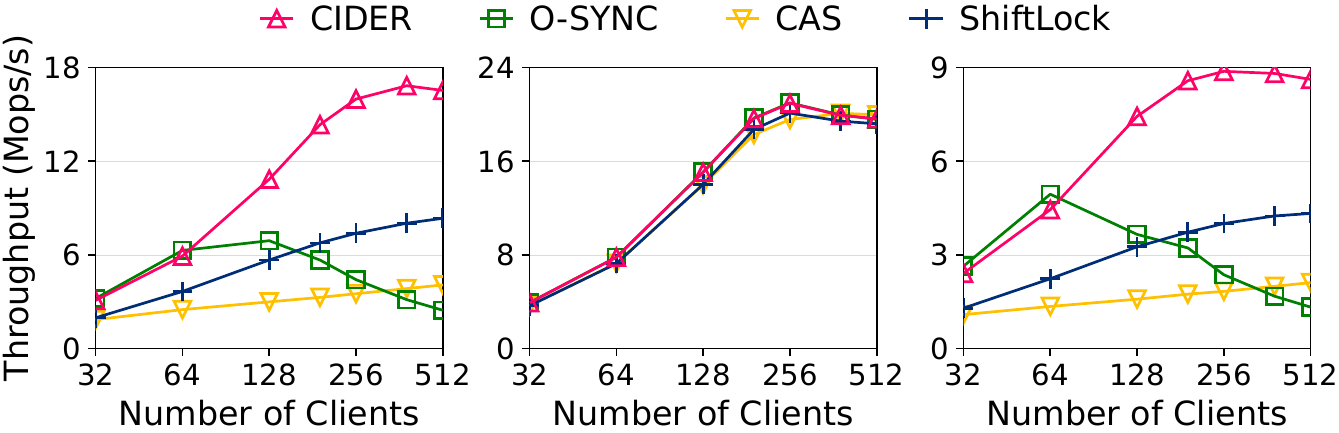}
        };
        \begin{scope}[x={(image.south east)}, y={(image.north west)}]
            \node[below=-1mm of image.south west, xshift=0.20\textwidth, 
                   anchor=north] {\small{\textbf{(a) Write-intensive}}};
            \phantomsubcaption
            \label{fig:fg11a}
            \node[below=-1mm of image.south west, xshift=0.53\textwidth, 
                   anchor=north] {\small{\textbf{(b) Read-intensive}}};
            \phantomsubcaption
            \label{fig:fg11b}
            \node[below=-1mm of image.south west, xshift=0.84\textwidth, anchor=north] {\small{\textbf{(c) Write-only}}};
            \phantomsubcaption
            \label{fig:fg11c}            
        \end{scope}
    \end{tikzpicture}
        \caption{The throughput comparison on a pointer array.}
    \end{minipage}
    \begin{minipage}[t]{0.51\textwidth}
        \centering
        \begin{tikzpicture}
        \node[anchor=south west, inner sep=0] (image) at (0,0) {
           \includegraphics[width=\columnwidth]{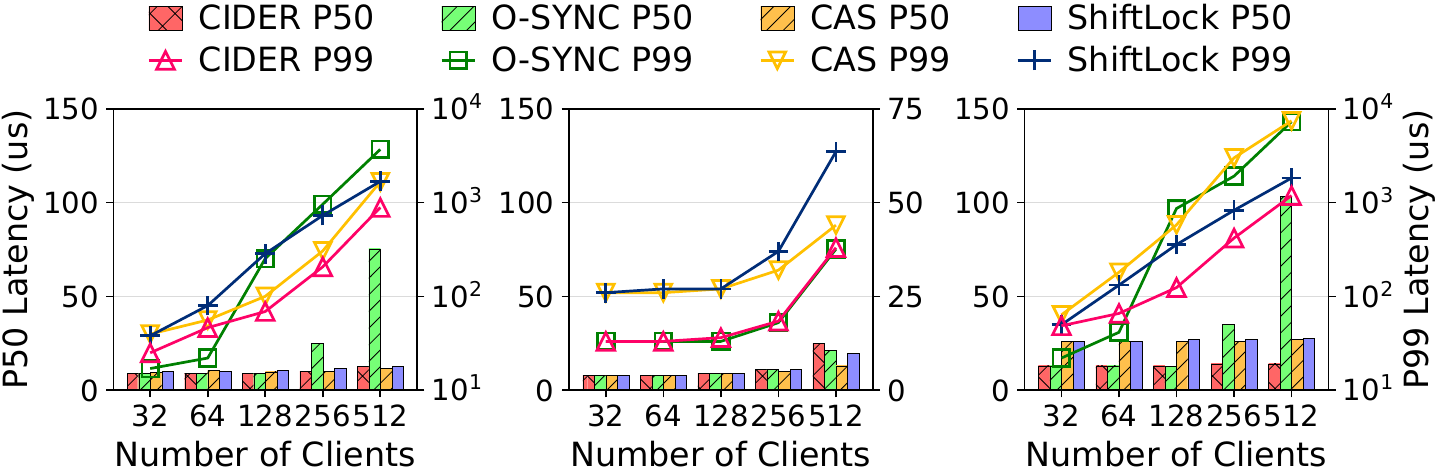}
        };
        \begin{scope}[x={(image.south east)}, y={(image.north west)}]
            \node[below=-1mm of image.south west, xshift=0.20\textwidth, 
                   anchor=north] {\small{\textbf{(a) Write-intensive}}};
            \phantomsubcaption
            \label{fig:fg12a}
            \node[below=-1mm of image.south west, xshift=0.51\textwidth, 
                   anchor=north] {\small{\textbf{(b) Read-intensive}}};
            \phantomsubcaption
            \label{fig:fg12b}
            \node[below=-1mm of image.south west, xshift=0.81\textwidth, 
                   anchor=north] {\small{\textbf{(c) Write-only}}};
            \phantomsubcaption
            \label{fig:fg12c}
        \end{scope}
        \end{tikzpicture}
        \caption{The latency comparison on a pointer array.}
    \end{minipage}
\end{figure*}

\begin{figure*}[t]
    \centering
    \begin{minipage}[t]{0.85\columnwidth}
        \centering
        \begin{tikzpicture}
        \node[anchor=south west, inner sep=0] (image) at (0,0) {
           \includegraphics[width=\columnwidth]{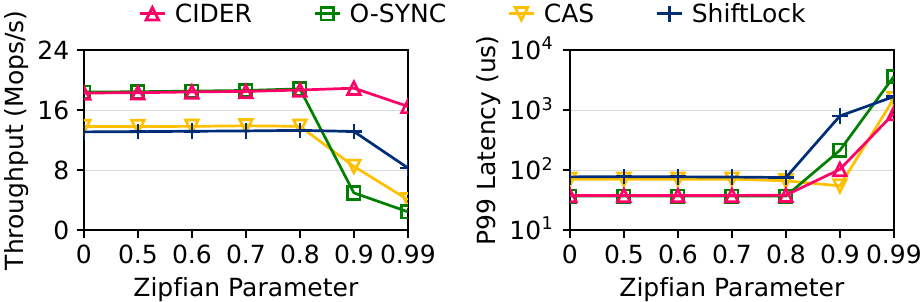}
        };
        \end{tikzpicture}
        \caption{The throughput and latency of \DMCS and baseline approaches as a function of the skewness factor (larger values mean more skew).}
        \label{fig:fg13}
    \end{minipage}
     \hspace{3mm}
    \begin{minipage}[t]{0.42\columnwidth}
      \centering
       \begin{tikzpicture}
        \node[anchor=south west, inner sep=0] (image) at (0,0) {
           \includegraphics[width=\columnwidth]{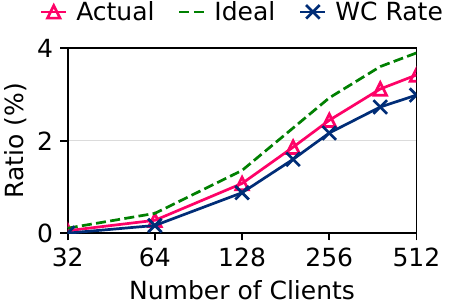}
        };
        \end{tikzpicture}
    \caption{The analysis of the ratio of pessimistic synchronization.}
    \label{fig:fg14}
    \end{minipage}
    \hspace{3mm}
        \begin{minipage}[t]{0.6\columnwidth}
                \centering
        \begin{tikzpicture}
        \node[anchor=south west, inner sep=0] (image) at (0,0) {
           \includegraphics[width=\columnwidth]{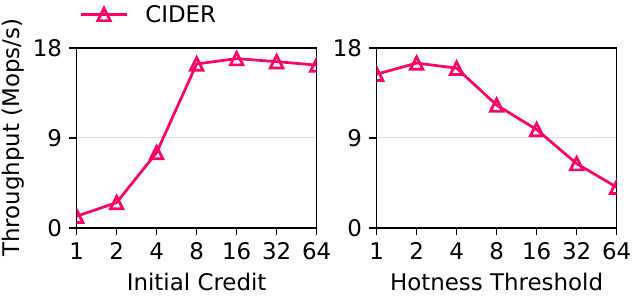}
        };
        \end{tikzpicture}
        \caption{The impact of contention-aware synchronization parameters on throughput.}
        \label{fig:fg15}
    \end{minipage}
\end{figure*}

\subsection{Experimental Setup}\label{sec:setup}
\noindent\textbf{Testbed.}
We run our experiments on 8 physical servers on CloudLab~\cite{duplyakin2019cloudlab}.
Each machine has two 36-core Intel Xeon CPUs, 256 GB DRAM, and a 100 Gbps Mellanox ConnectX-6 NIC.
All machines are connected by a 100 Gbps Ethernet switch.
Following previous work~\cite{wang2022sherman,luo2023smart,luo2024chime}, we configure one machine to serve as both CN and MN to save machine resources,
aligning with previous work~\cite{luo2023smart,shen2023ditto} in the CN-MN ratio.
Since we only have access to a limited number of physical servers, we emulate a large-scale cluster environment with up to 128 CNs by assigning 4 physical CPU cores to each CN, with each core running an independent client.

\noindent\textbf{\textit{Workloads.}}
We use YCSB workloads~\cite{cooper2010benchmarking} with 8-byte keys and 8-byte values~\cite{wang2022sherman,luo2023smart,luo2024chime,li2023rolex}, as it effectively captures the highly skewed and concurrent workloads characteristic of memory-disaggregated KV stores.
Similar to previous work~\cite{wang2022sherman, an2023marlin, wang2025dmtree}, we generate three types of workloads, \ie write-intensive, read-intensive, and write-only, where write operations include updating existing keys or inserting a new key if it does not exist.
The write-intensive workload is consistent with that described in Section~\ref{sec:background-and-motivations} and ~\ref{sec:analysis}.
We did not include range query workloads since KV stores typically do not support range queries (e.g., when using hashing).
The detailed workload descriptions are listed in Table~\ref{tab:workload}.
Unless otherwise specified, all workloads follow the Zipfian distribution with a skewness parameter of 0.99, which is representative of real-world workloads~\cite{cooper2010benchmarking}.

\noindent\textbf{\textit{Baselines.}}
We compare \DMCS with three synchronization schemes to show the efficacy of \DMCS.
We apply local WC~\cite{luo2023smart} to all baselines to achieve better performance.
\begin{itemize}[noitemsep, topsep=0pt, parsep=0pt, partopsep=0pt]
    \item Optimistic synchronization (O-SYNC): O-SYNC atomically modifies data pointers after writing the new KV data.
    \item CAS: Synchronizing data pointer modifications with the lock proposed by the \textsc{Smart}-framework~\cite{ren2024smart}, \ie a spinlock built on RDMA\_CAS with truncated exponential backoff for contention mitigation.
    \item ShiftLock: Use ShiftLock~\cite{gao2025shiftlock} to synchronize data pointer modifications. We disable the reader-writer lock due to its poor performance under KV workloads.
\end{itemize}

\noindent\textbf{\textit{Applications.}}
We evaluate \DMCS and the baselines on three KV stores with different index structures, \ie a pointer array, \textit{RACE}~\cite{zuo2021race}, and \textit{SMART}~\cite{luo2023smart}, respectively.
The pointer array is used to show the pure performance of data pointer modifications, where each pointer corresponds to an individual key.
\textit{RACE} and \textit{SMART} are two memory-disaggregated KV stores, which index KV pairs using the hash table and the radix tree, respectively.
They adopt the optimistic synchronization with out-of-place KV modification to support variable-length keys and values, thus benefiting from the \DMCS design.
For each application, we populate 60 million KV items before the evaluation and use their default configurations.

\subsection{Micro-Benchmarks}\label{sec:microbench}

We first use the pointer array application as a micro-benchmark to evaluate the pure performance gain of \DMCS's design.

\subsubsection{Overall performance}
Figures~\ref{fig:fg11a} and \ref{fig:fg12a} show the throughput and latency of the pointer array under the write-intensive workload.
Under low concurrency with fewer than 64 clients, O-SYNC performs the best since only a few retries are generated due to conflicting IDU operations.
In this case, \DMCS can achieve comparable performance with O-SYNC since contention-aware synchronization switches \DMCS to the optimistic mode.
However, under high concurrency with more than 64 clients, O-SYNC experiences a dramatic performance collapse due to the severe I/O redundancy of optimistic synchronization.
In this case, \DMCS scales well and achieves up to $6.7\times$ higher throughput and $4.2\times$ lower P99 latency compared with O-SYNC.
This is because \DMCS switches into the pessimistic mode, where the MCS lock reduces redundant I/O by queuing conflicting IDU operations in a global wait queue and performing the operations sequentially.
Furthermore, \DMCS outperforms ShiftLock by up to $2.0\times$ in throughput and $1.9\times$ in P99 latency since global WC further reduces the inter-node redundant I/Os, which achieves a higher combining efficiency than the local WC design.
Finally, CAS outperforms O-SYNC when the number of clients exceeds 384, indicating the inefficiency of optimistic synchronization under high-concurrency workloads.
However, it is inferior to ShiftLock and \DMCS because it still incurs I/O redundancy due to unsuccessful retries.

Figures~\ref{fig:fg11b} and \ref{fig:fg12b} show the throughput and latency under the read-intensive workload.
All baselines and \DMCS exhibit a similar performance in throughput. 
\DMCS's latency is comparable to O-SYNC, as both approaches utilize optimistic synchronization without additional lock maintenance overhead. 
In contrast, ShiftLock and CAS incur two additional RTTs to 5\% of write requests and thus increase the P99 latency by up to $2.1\times$.
The results verify the effect of the contention-aware synchronization of \DMCS.

Figures~\ref{fig:fg11c} and \ref{fig:fg12c} show the throughput and tail latency under the write-only workload.
The results are similar, where \DMCS achieves $6.5\times$, $4.1\times$, $2.0\times$ higher throughput and $6.1\times$, $6.3\times$, $1.5\times$ lower P99 latency compared with O-SYNC, CAS and ShiftLock.

As shown in Figure~\ref{fig:fg13}, we evaluate how workload skewness affects the performance of the pointer array under the write-intensive workload.
\DMCS performs best under both the uniform workload and highly skewed workloads. O-SYNC shows a good performance in the uniform workload while having the poorest performance in highly skewed workloads, \ie when the skewness is larger than 0.8. This is because the I/O redundancy issue becomes more severe under highly skewed workloads. ShiftLock and CAS perform better than O-SYNC in highly skewed workloads, since the pessimistic synchronization avoids I/O retries. However, they perform worse in the uniform workload due to the overhead of lock operations.

Figure~\ref{fig:fg14} quantifies the ratio of requests using pessimistic synchronization under write-intensive workloads.
Ideally, requests with severe contention conflicts, \ie requests whose retry count exceeds \texttt{HOTNESS\_THRESHOLD}, should adopt pessimistic synchronization. We define the proportion of such requests as the ideal pessimistic ratio, which is 4\% with 512 clients. Among these requests, \DMCS can accurately identify 88\% as suitable for pessimistic synchronization. Among requests using pessimistic synchronization, 87\% are combined, effectively reducing redundant operations.


\subsubsection{Parameter Selection}
Through experiments with 512 clients under our write-intensive workload, we justify our parameter selection, \ie \texttt{INITIAL\_CREDIT} as $36$ and \texttt{HOTNESS\_THRESHOLD} as 2. 
The left part of Figure~\ref{fig:fg15}  indicates that a small \texttt{INITIAL\_CREDIT} causes the transition to code keys to become too sensitive. 
Setting the values $>8$ stabilizes the throughput. 
The right part of Figure~\ref{fig:fg15} demonstrates that a higher \texttt{HOTNESS\_THRESHOLD} imposes stricter criteria for the transition to hot keys, shifting system behavior closer to optimistic synchronization. 
Setting the threshold to 2 achieves the optimal.

\begin{figure*}[!t]
    \begin{minipage}[t]{0.47\textwidth}
        \centering
    \begin{tikzpicture}
        \node[anchor=south west, inner sep=0] (image) at (0,0) {
            \includegraphics[width=\columnwidth]{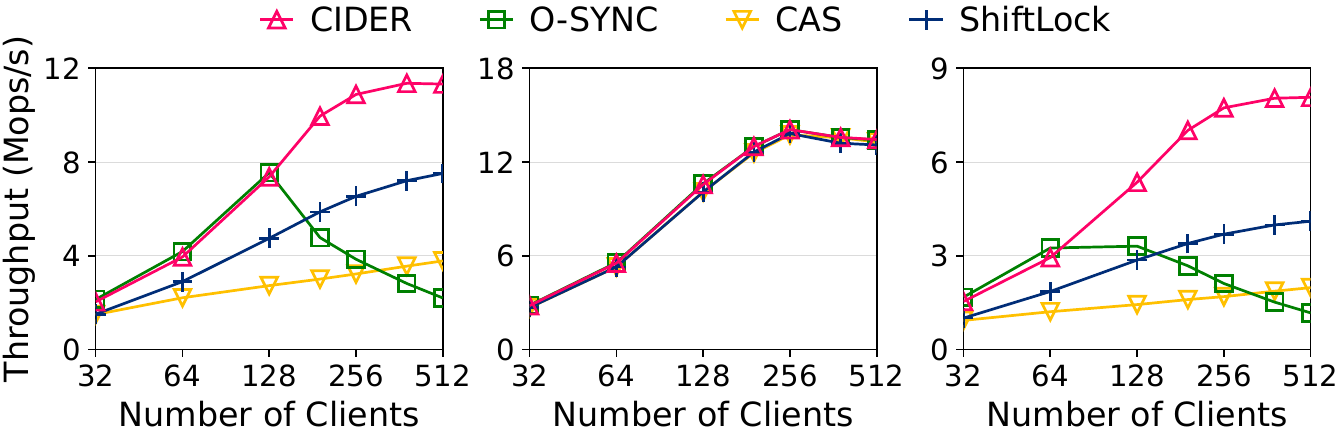}
        };
        \begin{scope}[x={(image.south east)}, y={(image.north west)}]
            \node[below=-1mm of image.south west, xshift=0.20\textwidth, 
                   anchor=north] {\small{\textbf{(a) Write-intensive}}};
            \phantomsubcaption
            \label{fig:fg16a}
            \node[below=-1mm of image.south west, xshift=0.53\textwidth, 
                   anchor=north] {\small{\textbf{(b) Read-intensive}}};
            \phantomsubcaption
            \label{fig:fg16b}
            \node[below=-1mm of image.south west, xshift=0.84\textwidth, anchor=north] {\small{\textbf{(c) Write-only}}};
            \phantomsubcaption
            \label{fig:fg16c}            
        \end{scope}
    \end{tikzpicture}
        \caption{The end-to-end throughput on \textit{RACE}.}
    \end{minipage}
    \begin{minipage}[t]{0.51\textwidth}
        \centering
        \begin{tikzpicture}
        \node[anchor=south west, inner sep=0] (image) at (0,0) {
           \includegraphics[width=\columnwidth]{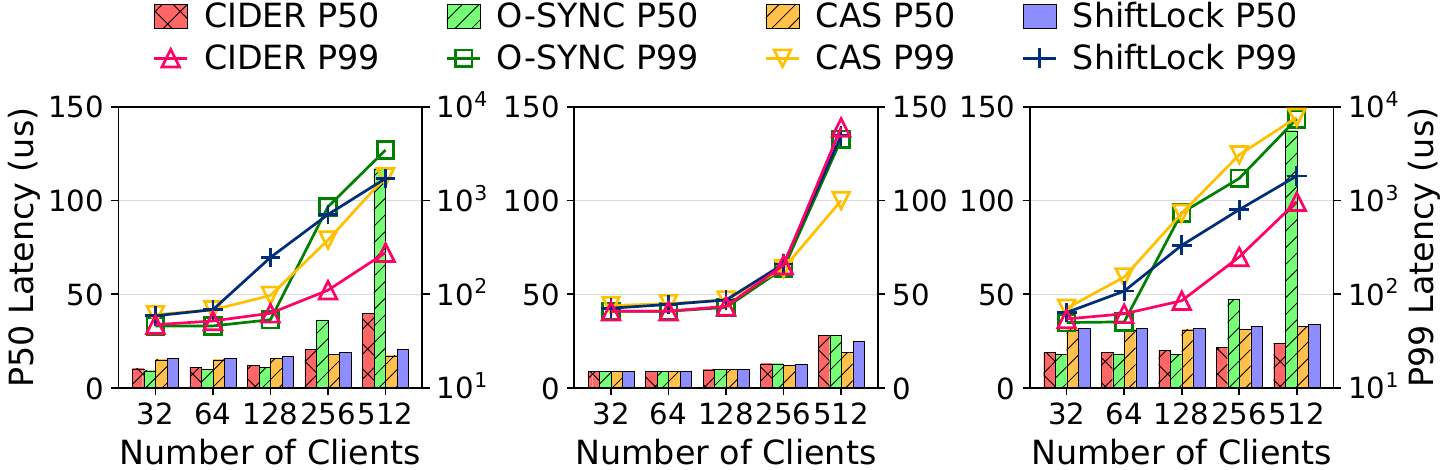}
        };
        \begin{scope}[x={(image.south east)}, y={(image.north west)}]
            \node[below=-1mm of image.south west, xshift=0.20\textwidth, 
                   anchor=north] {\small{\textbf{(a) Write-intensive}}};
            \phantomsubcaption
            \label{fig:fg17a}
            \node[below=-1mm of image.south west, xshift=0.51\textwidth, 
                   anchor=north] {\small{\textbf{(b) Read-intensive}}};
            \phantomsubcaption
            \label{fig:fg17b}
            \node[below=-1mm of image.south west, xshift=0.81\textwidth, 
                   anchor=north] {\small{\textbf{(c) Write-only}}};
            \phantomsubcaption
            \label{fig:fg17c}
        \end{scope}
        \end{tikzpicture}
        \caption{The end-to-end latency on \textit{RACE}.}
    \end{minipage}
\end{figure*}

\begin{figure*}[!t]
    \begin{minipage}[t]{0.47\textwidth}
        \centering
    \begin{tikzpicture}
        \node[anchor=south west, inner sep=0] (image) at (0,0) {
            \includegraphics[width=\columnwidth]{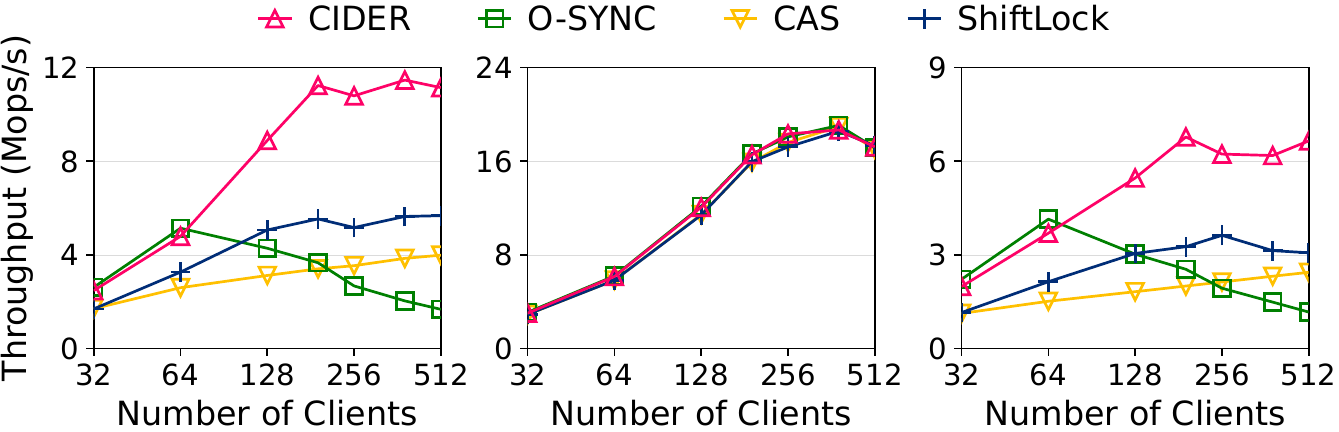}
        };
        \begin{scope}[x={(image.south east)}, y={(image.north west)}]
            \node[below=-1mm of image.south west, xshift=0.20\textwidth, 
                   anchor=north] {\small{\textbf{(a) Write-intensive}}};
            \phantomsubcaption
            \label{fig:fg18a}
            \node[below=-1mm of image.south west, xshift=0.53\textwidth, 
                   anchor=north] {\small{\textbf{(b) Read-intensive}}};
            \phantomsubcaption
            \label{fig:fg18b}
            \node[below=-1mm of image.south west, xshift=0.84\textwidth, anchor=north] {\small{\textbf{(c) Write-only}}};
            \phantomsubcaption
            \label{fig:fg18c}            
        \end{scope}
    \end{tikzpicture}
        \caption{The end-to-end throughput on \textit{SMART}.}
    \end{minipage}
    \begin{minipage}[t]{0.51\textwidth}
        \centering
        \begin{tikzpicture}
        \node[anchor=south west, inner sep=0] (image) at (0,0) {
           \includegraphics[width=\columnwidth]{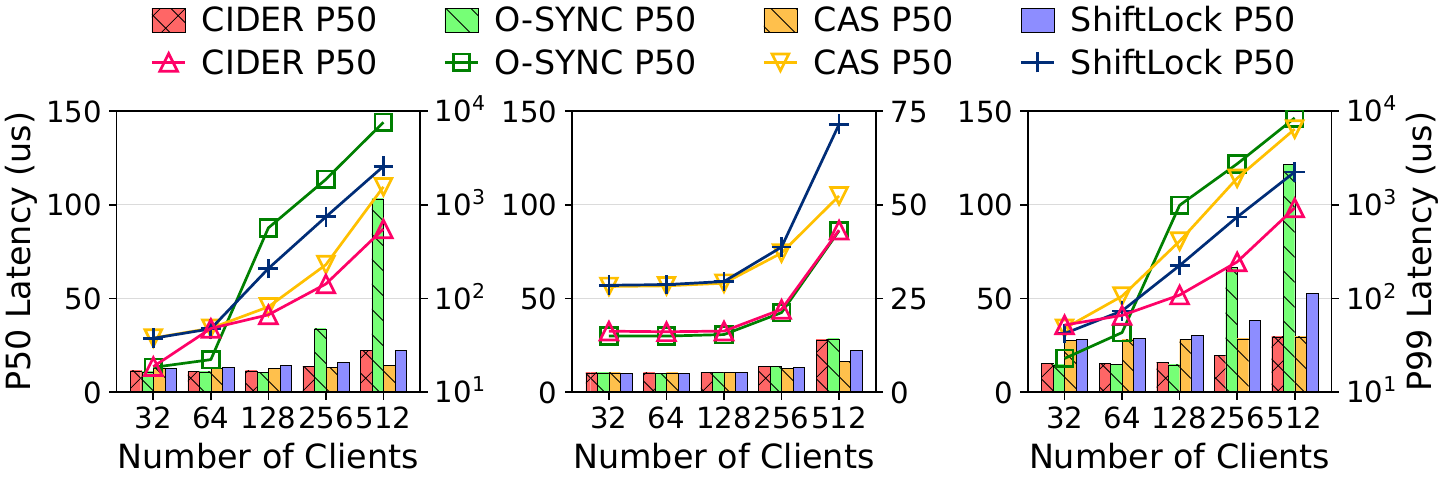}
        };
        \begin{scope}[x={(image.south east)}, y={(image.north west)}]
            \node[below=-1mm of image.south west, xshift=0.20\textwidth, 
                   anchor=north] {\small{\textbf{(a) Write-intensive}}};
            \phantomsubcaption
            \label{fig:fg19a}
            \node[below=-1mm of image.south west, xshift=0.51\textwidth, 
                   anchor=north] {\small{\textbf{(b) Read-intensive}}};
            \phantomsubcaption
            \label{fig:fg19b}
            \node[below=-1mm of image.south west, xshift=0.81\textwidth, 
                   anchor=north] {\small{\textbf{(c) Write-only}}};
            \phantomsubcaption
            \label{fig:fg19c}
        \end{scope}
        \end{tikzpicture}
        \caption{The end-to-end latency on \textit{SMART}.}
    \end{minipage}
\end{figure*}

\subsection{End-to-End Evaluation}\label{sec:e2eeval}


\subsubsection{RACE}
\textit{RACE}~\cite{zuo2021race} is a memory-disaggregated KV store that indexes KV pairs with a lock-free hash table.
\textit{RACE} employs optimistic synchronization by using RDMA\_CAS to modify data pointers. 
We integrate \DMCS into \textit{RACE} by first associating each slot entry with a lock entry to enable \textit{RACE} to support pessimistic synchronization.

\textbf{\textit{Performance under write-related workloads.}}
Figures~\ref{fig:fg16a} and \ref{fig:fg17a} show the throughput and latency of \textit{RACE} under the write-intensive workload.
Compared with O-SYNC, CAS and ShiftLock, \DMCS brings $5.1\times$, $3.0\times$, $1.5\times$ higher throughput and $12.4\times$, $6.4\times$, $6.2\times$ lower P99 latency to \textit{RACE}, respectively.
Even with local WC to reduce redundant I/Os, RACE still exhibits throughput degradation beyond 128 clients and performs worse under high contention.
This limitation stems from cross-CN concurrent requests that generate excessive redundant retry I/O, which cannot be eliminated by local WC.
The throughput improvement of \DMCS is less significant for \textit{RACE} ($5.1\times$) compared with that of the pointer array ($6.7\times$).
This is because \textit{RACE} requires additional RDMA\_READs to fetch remote hash buckets from the MN, making it bandwidth-bound.
Compared with ShiftLock, \DMCS exhibits a $1.9\times$ increase in P50 latency with 512 clients, because requests using optimistic synchronization experience higher contention.

Figures~\ref{fig:fg16c} and \ref{fig:fg17c} show the results under the write-only workload.
Compared with O-SYNC, CAS and ShiftLock, \DMCS brings $6.9\times$, $4.1\times$, $2.0\times$ higher throughput and $7.6\times$, $7.9\times$, $1.9\times$ lower P99 latency to \textit{RACE}.
Besides, compared with ShiftLock, \DMCS achieves a $1.9\times$ reduction in P99 latency on \textit{RACE}, which is less than the reduction under the write-intensive workload ($6.2\times$).
This is because the write-only workload generates higher concurrency, resulting in a longer wait queue in the global WC as well as more RTTs to unlock participants along the queue.

\textbf{\textit{Performance under the read-intensive workload.}}
All methods exhibit comparable throughput and latency, as seen in Figures~\ref{fig:fg16b} and \ref{fig:fg17b}. For O-SYNC, the 5\% write ratio is insufficient to cause an IOPS bottleneck. 
For CAS and Shiftlock, RACE’s two-choice design introduces higher read overhead across the 95\% reads, masking the lock overhead from the 5\% writes.

\subsubsection{SMART}
We also integrate \DMCS into \textit{SMART}~\cite{luo2023smart}, an advanced KV store on DM that uses the adaptive radix tree to index KV pairs.
To support variable-length KV data, \textit{SMART} adopts optimistic synchronization with out-of-place updates.

\textbf{\textit{Performance under write-related workloads.}}
Figures~\ref{fig:fg18a} and \ref{fig:fg19a} show the performance of \textit{SMART} with \DMCS under the write-intensive workload.
With O-SYNC, \textit{SMART} experiences a performance collapse with more than 128 clients due to the redundant retries on modifying data pointers.
On the contrary, \textit{SMART} indexes with CAS Lock, ShiftLock, and \DMCS scale well as the number of clients grows.
Compared with O-SYNC, CAS, and ShiftLock, \DMCS brings $6.6\times$, $2.8\times$, $2.0\times$ higher throughput and $13.8\times$, $2.8\times$, $4.7\times$ lower P99 latency to \textit{SMART}, respectively, since \DMCS reduces more redundant I/Os through global WC rather than local WC.
Figures~\ref{fig:fg18c} and \ref{fig:fg19c} show the performance under the write-only workload.
The evaluation results are similar, where \DMCS outperforms O-SYNC, CAS and ShiftLock by up to $5.7\times$, $2.7\times$,  and $2.2\times$ in throughput, respectively.

The throughput gains brought by \DMCS under write-only workloads are lower for \textit{SMART} ($5.7\times$) than for the pointer array ($6.5\times$).
This is because local WC provides more benefits in tree indexes than in the pointer array. 
Specifically, local WC can reduce intra-node redundant I/Os not only for data modifications but also for tree traversals.
Since the global WC relies on the MCS lock, it can only reduce redundant I/Os on the memory region protected by the lock.
Thus, it cannot combine tree traversals like the local WC.
Additional tree traversal occurs during write operations, consequently imposing more significant performance overhead under write-only workloads.
However, \DMCS still performs the best compared with baselines since its capability of reducing inter-node I/O redundancy.

\textbf{\textit{Performance under the read-intensive workload.}}
Figures~\ref{fig:fg18b} and \ref{fig:fg19b} show the performance of \textit{SMART} under the read-intensive workload.
The results are similar to those on a pointer array.

\subsection{Factor Analysis for \DMCS Design}

Finally, we investigate how our design improves performance in terms of throughput, latency, and the internal WC efficiency.

\begin{figure}[!t]
    \centering
    \includegraphics[width=1\linewidth]{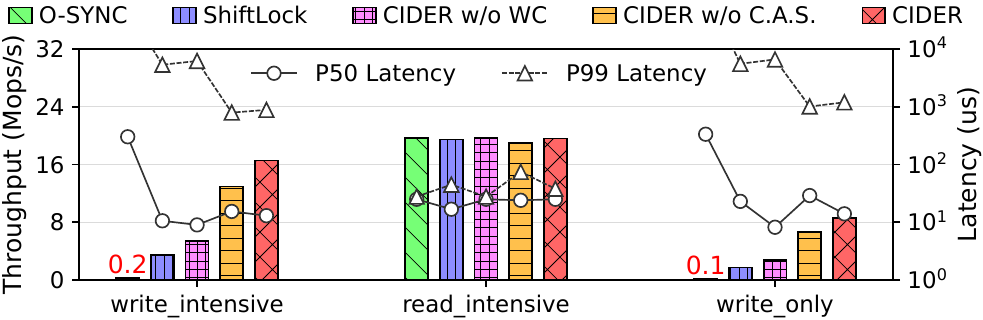}
    \caption{The factor analysis of overall performance on \DMCS. "C.A.S." represents contention-aware synchronization.}
    \label{fig:fg20}
\end{figure}

\begin{figure}[t]
    \centering
     \begin{minipage}[t]{0.5\columnwidth}
    \includegraphics[width=1\linewidth]{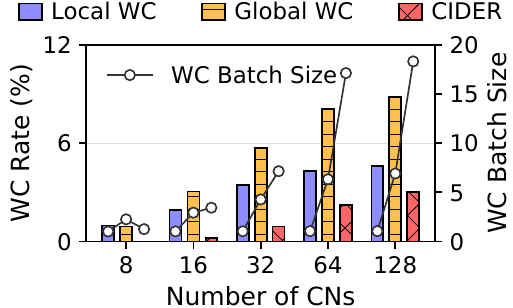}
    \caption{The efficiency comparison of different WC mechanisms.}\label{fig:fg21}
    \end{minipage}
    \hspace{3mm}
    \begin{minipage}[t]{0.43\columnwidth}
    \includegraphics[width=1\linewidth]{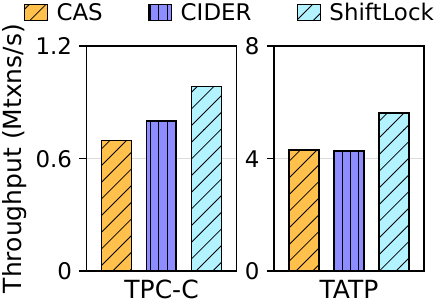}
    \caption{Performance under transactions.}\label{fig:fg22}
    \end{minipage}
\end{figure}

\subsubsection{Throughput and latency}
Figure~\ref{fig:fg20} presents the factor analysis for the techniques in \DMCS.
We disable local WC for both O-SYNC and ShiftLock to show the performance gain purely introduced by global WC.
The experiment is conducted on a pointer array using 512 clients.
Without loss of generality, we discuss the performance improvement under the write-intensive workload.

\textbf{\textit{+ Contention-aware synchronization (\DMCS w/o WC).}}
Compared with O-SYNC, contention-aware synchronization achieves a $22\times$ improvement in throughput as it prevents redundant retries on hot keys.
Compared with ShiftLock, we achieve $1.6\times$ higher throughput, owing to reduced operational overhead on cold keys. However, since the P99 latency is dominated by highly contended hot keys, no significant latency optimization is observed.


\textbf{\textit{+ Global write combining (\DMCS w/o C.A.S.).}}
Compared with ShiftLock, the global WC improves throughput by $3.7\times$ and reduces P99 latency by at least $5.4\times$ since it significantly reduces redundant I/Os by combining \texttt{UPDATE} operations, alleviating IOPS bottleneck on the MN.
The P50 latency remains unaffected, as the majority of requests do not induce severe I/O redundancy and therefore do not benefit from the global WC technique.

\textbf{\textit{\DMCS.}}
Integrating both global WC and contention-aware synchronization, \DMCS demonstrates superior performance compared with using either technique in isolation.



\subsubsection{WC efficiency}
We evaluate the \textit{WC rate} and the \textit{WC batch size} of local WC, global WC, and \DMCS in Figure~\ref{fig:fg21}.
The \textit{WC rate} is defined in Section~\ref{sec:backgound-pessimistic}, \ie the proportion of IDU operations that are combined by WC techniques.
The \textit{WC batch size} is defined as the average number of requests in each combined batch.

In terms of the WC rate, global WC has a $1.9\times$ higher WC rate than local WC, demonstrating a higher upper bound for combining efficiency.
However, despite having the highest WC rate, the throughput of global WC is still less than \DMCS since its small average WC batch size makes the communication overhead of conducting WC higher for clients.
\DMCS can achieve a larger average WC batch size thanks to the contention-aware synchronization. 
Specifically, \DMCS employs optimistic synchronization for cold keys to minimize locking overhead, while preserving global WC batching for hot keys, where the benefits are more substantial, improving the actual combining efficiency.


\subsection {Distributed Transactions}

We evaluate the performance of \DMCS in a transactional system using the TPC-C~\cite{tpcc} and TATP~\cite{tatp} benchmarks. Our setup follows the same parameter configurations as ShiftLock~\cite{gao2025shiftlock}. Specifically, we use six servers, \ie one lock server and five client servers, each hosting 64 clients. The system implements the typical two-phase locking (2PL) protocol, where clients acquire all required locks before execution and release them afterward. Clients simulate transaction execution via busy-waiting.

We disable \DMCS's contention-aware synchronization and the WC techniques as they violate the 2PL and transaction atomicity, respectively. We compare against two baselines: CAS lock and ShiftLock~\cite{gao2025shiftlock}, which implements reader-writer locks.

The left part of Figure~\ref{fig:fg22} shows TPC-C results. Due to the adoption of the MCS lock, \DMCS reduces redundant I/O compared with CAS lock, achieving $1.1\times$ higher throughput. ShiftLock achieves $1.2\times$ higher throughput than \DMCS since its reader-writer lock design supports concurrent reads. The right part of Figure~\ref{fig:fg22} presents TATP results. With lower contention, CAS lock and \DMCS perform similarly. ShiftLock achieves $1.3\times$ higher throughput due to its reader-writer lock design.

\section{Related Work}

\noindent
\textbf{\textit{Disaggregated Memory.}}
DM is a next-generation data center architecture that is widely discussed in both academia and industry.
Existing works on DM can be categorized into DM systems and DM applications.
\textit{DM systems} focus on achieving high-performance and transparent execution of applications on memory-disaggregated data centers.
Existing works span multiple levels, including hardware design~\cite{guo2022clio,wang2021concordia,lee2021mind}, operating systems~\cite{amaro2020fastswap,shan2018legoos,Vilanova2022fractos,zhang2022teleport, bergman2022osdm}, and software runtimes~\cite{Ruan2020aifm,wang2020semeru,wang2022memliner,chen2024tale2,Ma2024drust,zahka2022runtime,calciu2021runtime}.
\textit{DM applications} refer to a bottom-up approach that builds native applications directly over memory-disaggregated data centers.
Many applications have been ported to DM, \eg cache systems~\cite{zhang2021redy,shen2023ditto}, transaction systems~\cite{zhang2022ford,zhang2024motor}, databases~\cite{jang2023cxlanns,zhang2020db}, KV stores~\cite{shen2023fusee,zuo2021race,wang2022sherman,luo2023smart,li2023rolex,luo2024chime,lee2023dinomo}.


The most closely related work is \textsc{Smart}-framework~\cite{ren2024smart}, an I/O optimization framework on DM. It  improves throughput via thread-aware resource allocation and reduces CAS retry overhead with adaptive backoff. Unlike it, \DMCS focuses on addressing the redundant I/Os incurred by optimistic synchronization with the global WC and contention-aware synchronization.

\noindent
\textbf{\textit{Memory-Disaggregated Key-Value Stores.}}
Network I/O is the key bottleneck for memory-disaggregated KV stores.
Existing approaches focus on improving the I/O efficiency of KV stores on DM in a bottom-up manner, \ie by tailoring data structures and algorithms to reduce I/O sizes and numbers between CN and MNs.
Specifically, RACE~\cite{zuo2021race} is an extendible hash table with a lock-free concurrency control scheme. 
SMART~\cite{luo2023smart} proposes a radix tree design to avoid the read amplifications of B+ trees.
It further presents the read-delegation and write-combining technique to reduce redundant I/Os on DM.
CHIME~\cite{luo2024chime} is a hybrid index combining B+ trees and hopscotch hashing to reduce read amplifications of B+ trees.
FUSEE~\cite{shen2023fusee} adopts a two-level memory management technique, reducing frequent remote allocation I/Os. 
Different from these approaches, \DMCS is a general optimization method that can be applied to a large body of index data structures.
All index data structures that employ optimistic synchronization with out-of-place data modification can be optimized with \DMCS.
Hence, \DMCS is orthogonal to these data structures and algorithm designs.

\noindent
\textbf{\textit{RDMA-Based Lock Management.}}
RDMA has attracted increasing research attention in terms of distributed lock management, which can be classified into two types, \ie centralized and decentralized lock management.
\textit{Centralized lock management}~\cite{zhang2024fisslock, yu2020netlock} relies on a central server for granting locks, which is unfriendly to DM due to the limited compute capability at the memory side.
\textit{Decentralized lock management}~\cite{yoon2018dslr, deculapalli2005queuelock, narracula2007queuelock, wei2015drtm,wei2018lock} leverages one-sided RDMA verbs to bypass the CPU bottleneck.


Existing memory-disaggregated KV stores typically conduct lock acquisitions via RDMA\_CAS with a fail-and-retry strategy~\cite{wang2022sherman, luo2023smart, luo2024chime, zhang2024motor, li2023rolex}.
The CAS-based lock will rapidly saturate the limited IOPS upper bound of memory-side RNICs, resulting in poor scalability. 
ShiftLock~\cite{gao2025shiftlock} proposes an RDMA-based MCS lock design to address this issue, which can also be applied to DM.
Different from existing approaches, \DMCS proposes a global WC design on top of the MCS lock to further reduce redundant data modifications.

\section{Conclusion}
In this paper, we identify that current memory-disaggregated KV stores face significant performance bottlenecks due to redundant I/O operations when synchronizing concurrent data accesses. 
This issue stems from a fundamental mismatch between their optimistic synchronization schemes and the highly contended workloads prevalent in these systems. 
To address this, we propose to adopt pessimistic synchronization strategies to enhance performance. 
Based on this idea, we design and implement \DMCS, a compute-side I/O optimization framework, with two key techniques, \ie global write combining and contention-aware synchronization. 
Our evaluation demonstrates that \DMCS significantly improves the throughput of leading memory-disaggregated storage systems by up to $6.6\times$ under the write-intensive workload.
\begin{acks}
We sincerely thank our anonymous shepherd and reviewers for helping us improve our paper.
This work is supported by the National Natural Science Foundation of China (Project No. 62472101) and the Open Fund of PDL (Project No.WDZC20245250106). Jiacheng Shen and Xin Wang are corresponding authors.
\end{acks}

\bibliographystyle{ACM-Reference-Format}
\bibliography{mcs.bib}

\onecolumn
\newpage
\twocolumn
\section*{APPENDIX}

\subsection*{A \hspace{6pt} SUPPLEMENTARY SENSITIVITY ANALYSIS}

\DMCS consistently maintains its optimization effectiveness across various scenarios (including different data scales and key-value sizes). We conduct the following sensitivity tests, highlighting its capability to enhance efficiency in diverse production environments.



\begin{figure}[t]
    \begin{minipage}[t]{0.47\textwidth}
        \centering
        \begin{tikzpicture}
        \node[anchor=south west, inner sep=0] (image) at (0,0) {
           \includegraphics[width=\columnwidth]{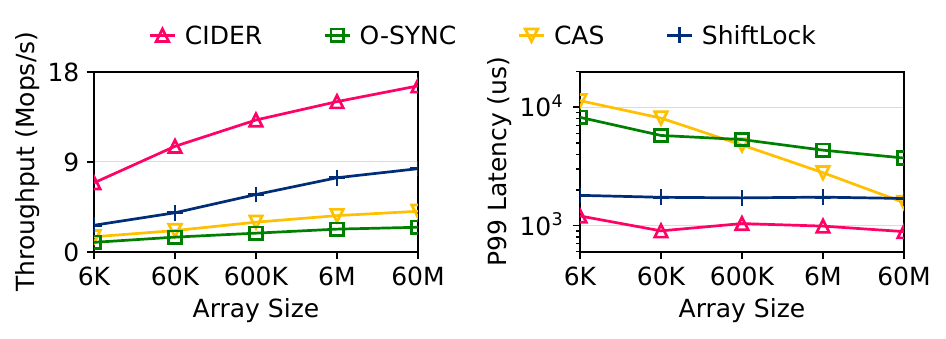}
        };
        \begin{scope}[x={(image.south east)}, y={(image.north west)}]
            \node[below=-1mm of image.south west, xshift=0.27\textwidth, 
                   anchor=north] {\small{\textbf{(a) The throughput.}}};
             \phantomsubcaption
            \label{fig:fg23a}
            \node[below=-1mm of image.south west, xshift=0.78\textwidth, 
                   anchor=north] {\small{\textbf{(b) The tail latency.}}};
             \phantomsubcaption
            \label{fig:fg23b}
        \end{scope}
        \end{tikzpicture}
        \caption{The performance comparison as a function of the array size.}
        \label{fig:fg23}
    \end{minipage}
\end{figure}

\textbf{\textit{The impact of array sizes.}}
As shown in Figure~\ref{fig:fg23}, we evaluate the impact of array size on a pointer array under the write-intensive workload.
When the array size is relatively small, limited workload concurrency becomes a severe performance bottleneck.
This is because a majority of clients are contending for the same entry, synchronizing either by the MCS wait queue or the atomic RDMA\_CAS.
As the array size increases, this contention is alleviated, leading to improved throughput across all methods.

\begin{figure}[t]
    \begin{minipage}[t]{0.47\textwidth}
        \centering
        \begin{tikzpicture}
        \node[anchor=south west, inner sep=0] (image) at (0,0) {
           \includegraphics[width=\columnwidth]{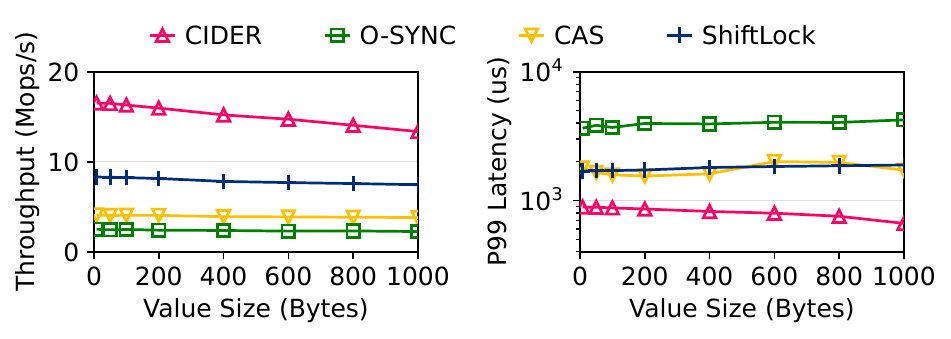}
        };
        \begin{scope}[x={(image.south east)}, y={(image.north west)}]
            \node[below=-1mm of image.south west, xshift=0.27\textwidth, 
                   anchor=north] {\small{\textbf{(a) The throughput.}}};
             \phantomsubcaption
            \label{fig:fg24a}
            \node[below=-1mm of image.south west, xshift=0.78\textwidth, 
                   anchor=north] {\small{\textbf{(b) The tail latency.}}};
             \phantomsubcaption
            \label{fig:fg24b}
        \end{scope}
        \end{tikzpicture}
        \caption{The performance comparison as a function of the value size.}
        \label{fig:fg24}
    \end{minipage}
\end{figure}

\textbf{\textit{The impact of value sizes.}}
As shown in Figure~\ref{fig:fg24}, we evaluate the impact of different value sizes on a pointer array under the write-intensive workload.
As the size of the value increases, the throughput of all baselines remains stable.
This is because \DMCS and all baselines are IOPS-bound rather than bandwidth-bound, as they employ either local or global WC techniques to eliminate redundant \texttt{update} operations, alleviating bandwidth pressure on the memory-side RNICs.





\end{document}